\documentclass[prb,twocolumn,showpacs,preprintnumbers,amsmath,amssymb]{revtex4}

\usepackage{dcolumn}
\usepackage{bm}
\usepackage{wrapfig}
\usepackage{subfig}

\usepackage{graphicx,color}

\renewcommand{\baselinestretch}{1.0}
\newcommand{\captionline}{\renewcommand{\baselinestretch}{1.0}}
\newcommand{\myfonts}{}

\newcommand\Dfrtl[1]{\ensuremath{\,\mathrm{d}#1\,}}
\newcommand\imag{i}
\renewcommand\Re{\ensuremath{\mathrm{Re}\,}}
\renewcommand\Im{\ensuremath{\mathrm{Im}\,}}

\newcommand\euler[1]{\ensuremath{\mathrm{e}^{#1}}}

\begin{document}

\title{Imaginary-time quantum many-body theory out of equilibrium I: Formal
equivalence to Keldysh real-time theory and calculation of static properties}

\date{\today}
\author{Jong E. Han}
\email{jonghan@buffalo.edu}
\affiliation{Department of Physics, State University of New York at
Buffalo, Buffalo, NY 14260, USA}
\author{Andreas Dirks, and Thomas Pruschke}
\affiliation{Department of Physics, University of G\"ottingen, D-37077
G\"ottingen, Germany}

\pacs{73.63.Kv, 72.10.Bg, 72.10.Di, 72.15.Qm}

\begin{abstract}
We discuss the formal relationship between the real-time Keldysh and
imaginary-time theory for nonequilibrium in quantum dot systems. The
latter can be reformulated using the recently proposed Matsubara voltage
approach. We establish general conditions for correct analytic
continuation procedure on physical observables, and apply the technique
to the calculation of static quantities in steady-state non-equilibrium
for a quantum dot subject to a finite bias voltage and external magnetic
field. Limitations of the Matsubara voltage approach are also pointed
out.
\end{abstract}

\maketitle

\section{Introduction}\label{sec:I}

Experimental investigation of solids is in most cases concerned with observation
of static or dynamic properties in a weakly perturbed macroscopic system. Therefore,
standard techniques from equilibrium statistical mechanics are usually sufficient,
possibly supplemented by linear-response theories to account for transport.
Equilibrium statistical mechanics is based on the Gibbsian approach where
the statistical density matrix of a state at energy $E_s$ is given by the Boltzmann
factor $e^{-\beta (\hat{H}-\mu\hat{N})}$ with inverse temperature $\beta=1/k_{\rm B}T$ 
and the chemical potential
$\mu$. The big success in the theoretical description of quantum systems in thermal
equilibrium is based on the fact that both the thermal average and time evolution are
based on the same operator, and one can use the concept of Wick rotation to
formulate a theory which actually condenses  both type of dynamics into a single
complex Matsubara frequency theory. 

The advances in experimental methods over the past two decades have however opened
the access to studies, where time dependencies on the scale of internal time-scales become
visible,\cite{kirchmann} or where mesoscopic systems can be driven out of thermal equilibrium in a controlled way
and various properties can be experimentally observed, both in steady-
and time-dependent states.
Therefore, one pressing question to modern quantum many-body theory is how one can describe
generic non-equilibrium situations in macroscopic or mesoscopic systems. For the latter the 
paradigms are the single-electron quantum dot and nano-wires, where a tremendous amount
of data on transport or transient response has been collected over the past ten years.\cite{datta_book,hanson}

Out-of-equilibrium many-body theory is an emerging field which poses an
extreme challenge.  There are many attempts to use existing theoretical
approaches, the most popular being the ones based on the Keldysh
formulation of perturbation theory.\cite{rammer} In particular, the
growing interest in transport through mesoscopic systems triggered a
variety of applications of this technique; for example direct
perturbation theory with respect to different zeroth order
Hamiltonians,\cite{hershfield,meir,ueda}  functional renormalization
group methods or their derivatives\cite{schoeller,rosch,gezzi} or direct
numerical evaluation of the real-time
propagators.\cite{andersschiller,thorwart,werner_rtqmc,boulat,feiguin}
There are many other ideas, for example based on the concept of
infinitesimal unitary transformations.\cite{hackl_et_al} A comprehensive
overview can for example be found in Ref.\ \onlinecite{eckstein_review}.

An early attempt to formulate an out-of-equilibrium version of statistical mechanics
for steady-state properties of general quantum many-body systems is due to Zubarev,\cite{zubarev} who  tried
to construct a time-independent density matrix formalism by solving the equation of
motion within the scattering state formalism. This approach has later been revisited
by Hershfield in the context of  transport through quantum dot systems.\cite{hershfield_Y} 
The main problem with these, in principle exact formulations, is that they cannot be readily
applied because they require the solution of the Lippmann-Schwinger equation for
the scattering states, which amounts to knowing the full solution itself. There have been
attempts to tackle this problem by utilizing advanced tools of quantum many-body theory like Bethe ansatz\cite{mehta} or an extension of Wilson's numerical renormalization techniques.\cite{anders}
However, the former approach could only be applied to a very specific model, while the latter 
may lack a thorough foundation regarding the proper steady-state limit.\cite{rosch_critics}

In the present manuscript we focus on a different way to
extend the theoretical framework of equilibrium quantum mechanics to steady-state
nonequilibrium for quantum impurity models via an imaginary-time theory. We especially discuss the
possibility to deform the complex time contour for physical observables in
equilibrium to the Keldysh contour appropriate
for nonequilibrium, as proposed by Doyon and Andrei~\cite{doyon}. One fundamental
problem that arises in any such attempt stems from the fact that the non-equilibrium
steady-state Boltzmann factor and the time-evolution operator now have a fundamentally
different structure, and thus a straight-forward Wick rotation is not possible. As an
alternative procedure, we show that, by introduction of
\textit{Matsubara voltage}, the problem of the dual operators can be
resolved and a consistent theory for steady-state non equilibrium based on auxiliary
statistical mechanical problems formulated.

As the first step we need to properly define in what sense we achieve a steady state in a
quantum impurity model. This is done in section \ref{sec:II} together with a discussion of the general
structure for Keldysh perturbation theory, the problem of analytical continuation and the
idea of the Matsubara voltage formulation. The equivalence of the Keldysh real-time
and the Matsubata voltage perturbation theory for the steady state will be shown in section \ref{sec:III}
for the single-impurity Anderson model. In section \ref{sec:IV} we derive expressions for
calculating static observables on the impurity via an analytical continuation procedure from
the Matsubara voltage description. As summary, section \ref{sec:V} will conclude the paper.

Since many details are rather technical and not really necessary to understand the main line
of argument, we included them in a series appendices, which will be referred to when necessary.

\section{Many-body theory off equilibrium}\label{sec:II}

\subsection{Convergence to steady-state nonequilibrium}\label{sec:converge}
To establish a steady-state nonequilibrium, one requires the system to
be in the infinite-size limit. In mesoscopic systems, such as quantum dots,
this requirement means that the size of the reservoirs,
$L$, should be the largest scale. In particular,  
the time $t_W$ for the wake of
the perturbation occurring in the quantum dot region to reach the edge of
the reservoir with the Fermi velocity $v_F$ ($t_W=L/v_F$) should be greater than
any time scale used for the turn-on of the perturbation or measurements.
This ensures that the reflected wave does not interfere with the
formation of the steady-state and its measurements.
Alternatively, the reciprocal $v_F/L$ also represents the level spacing of the
continuum states, which sets the smallest energy scale in the model.

As in conventional many-body theory, we consider a perturbation which we turn on infinitesimally slow with a 
rate $\eta^{-1}$ as
\begin{equation}
\hat{V}(t)=\hat{V}e^{\eta t}
\label{eq:vt}
\end{equation}
for the time interval $t\in [-T,0]$, where $T$ is some initial time which eventually will be sent to infinity.
For $t>0$, the perturbation remains
constant at the full strength, $\hat{V}(t)=\hat{V}$. The above
discussions lead to the
relation between the three energy scales
\begin{equation}
\frac{v_F}{L}\ll \frac{1}{T} \ll \eta.
\end{equation}
In his original proposal\cite{hershfield}, Hershfield assumed the presence of an external
relaxation process to derive the time-independent density matrix in the limit $T\to\infty$. 
Recently Doyon
and Andrei\cite{doyon} have shown that for mesoscopic systems infinite reservoirs provide a relaxation process and any
assumption of an additional external relaxation source is not necessary. 
This suggests that we can do away
with the adiabatic factor $e^{\eta t}$ in a time-dependent theory as
long as the limit $L\to\infty$ is taken first.
Here we show through an explicit calculation that the adiabatic factor 
$e^{\eta t}$ is not necessary for the steady-state if local measurements
are made near the quantum dot\cite{taylor},
henceforth abbreviated as QD.

Our model system consists of a QD connected to two
fermionic reservoirs labeled by $\alpha=L,R$ (or
$\pm1$, respectively, when the reservoir index is taken numerically). We
include the single-particle tunneling between the leads and the QD into the non-interacting part of the Hamiltonian, which then becomes the resonant level model (RLM)
\begin{eqnarray}
\hat{H}_0 & = & \sum_{\alpha k\sigma}\epsilon_{\alpha k\sigma}c^\dagger_{\alpha k\sigma}
c_{\alpha k\sigma}+\epsilon_d\sum_\sigma d^\dagger_\sigma d_\sigma
\nonumber \\
& & -\sum_{\alpha k\sigma}\frac{t_\alpha}{\sqrt\Omega}(d^\dagger_\sigma
c_{\alpha k\sigma}+\text{h.c.})\;\;.
\end{eqnarray}
Here, $c^\dagger_{\alpha k\sigma}$ is the creation operator of conduction
electrons for the reservoir $\alpha$ with energy $\epsilon_{\alpha
k\sigma}$ at the continuum index $k$ and spin $\sigma$; $d^\dagger_\sigma$ creates an electron
on the QD orbital and $t_\alpha$ is the tunneling integral. $\Omega$ is
the normalization due to the volume of the reservoirs. This Hamiltonian can
be diagonalized by the scattering state operators $\psi^\dagger_{\alpha
k\sigma}$ given by the formal Lippmann-Schwinger operator equation,
\begin{equation}
\psi^\dagger_{\alpha k\sigma}
=c^\dagger_{\alpha k\sigma}
-\frac{t_\alpha}{\sqrt\Omega}\frac{1}{\epsilon_{\alpha k\sigma}-{\cal
L}_0+i0^+}d^\dagger_\sigma,
\end{equation}
with the Liouville operator acting on the operator space as
${\cal L}_0{\cal O}=[\hat{H}_0,{\cal O}]$. 
This equation can be easily solved as
\begin{eqnarray}
\psi^\dagger_{\alpha k\sigma}
&=&c^\dagger_{\alpha
k\sigma}-\frac{t_\alpha}{\sqrt\Omega}g_0(\epsilon_{\alpha k\sigma})d^\dagger
\nonumber\\
&+&\sum_{\alpha' k'\sigma}\frac{t_\alpha
t_{\alpha'}}{\Omega}\frac{g_0(\epsilon_{\alpha k\sigma})
c^\dagger_{\alpha' k'\sigma}}{\epsilon_{\alpha
k\sigma}-\epsilon_{\alpha' k'\sigma}+i0^+},
\end{eqnarray}
with the bare retarded Green's function for the QD, $g_0(\omega)
=(\omega-\epsilon_d+i\Gamma)^{-1}$. Here, 
$\Gamma=\pi (t_L^2+t_R^2)N(0)$ is the hybridization broadening, and we assume  
for simplicity a flat  density of states $N(0)$ for both reservoirs.

According to Hershfield~\cite{hershfield}, the
nonequilibrium steady-state created by a shift of chemical potential on
the source (drain) reservoir by $\Phi/2$ ($-\Phi/2$) can be described by
a density matrix
\begin{equation}
\hat\rho_0=\frac{\exp[-\beta(\hat{H}_0-\Phi\hat{Y}_0)]}{\Xi},
\end{equation}
with the so-called $Y$-operator defined as
\begin{equation}
\hat{Y}_0=\frac12\sum_{k\sigma}\left(
\psi^\dagger_{Lk\sigma}\psi_{Lk\sigma}
-\psi^\dagger_{Rk\sigma}\psi_{Rk\sigma}
\right)
\end{equation}
and the generalized partition function
$$
\Xi=\text{Tr}\exp[-\beta(\hat{H}_0-\Phi\hat{Y}_0)]\;\;.
$$
Since $\hat{Y}_0$ is diagonal in the eigen-operators,
$[\hat{H}_0,\hat{Y}_0]=0$ and $\hat\rho_0$ is time-independent.
It is important to realize that the convergence factor $i0^+$ in the
denominator of the Lippmann-Schwinger equation determines that the one
particle states $c^\dagger_{\alpha k\sigma}$ originate from the infinite
past inside the reservoir of infinite size. Thus the limit $L\to\infty$ has
already been taken implicitly before the perturbation is turned on.

\subsection{Real-time theory for open system}
\label{sec:real}

In addition to the noninteracting part $H_0$, the full Hamiltonian $H$
of the system will in general also contain an interaction we will denote
as $\hat{V}$ in the following.
For a general observable $\hat{A}$, we define its nonequilibrium
expectation value as
\begin{equation}
\lim_{T\to \infty}\langle \hat{A}(T)\rangle
=\lim_{T\to\infty}
\frac{{\rm
Tr}\left(e^{i\hat{H}T}\hat{A}e^{-i\hat{H}T}\hat\rho_0\right)}{
{\rm Tr}\hat\rho_0},
\end{equation}
where $\hat{A}$
has been evolved with the full Hamiltonian $\hat{H}$ during the time
interval $-T<t<0$. 
{Unlike Eq.~(\ref{eq:vt}), here we
take $\hat{V}(t)=\hat{V}$ for $-T<t<0$.} Defining the time-dependent
operator $\hat{A}(t)$ in the Heisenberg picture, $\hat{A}(t)
=e^{i\hat{H}t}\hat{A}e^{-i\hat{H}t}$, $\hat{A}(t)$ satisfies
$\frac{d}{dt}\hat{A}(t)=i[\hat{H},\hat{A}(t)]$ and
\begin{equation}
\hat{A}(t)=\hat{A}+i\int_0^t dt'[\hat{H},\hat{A}(t')].
\end{equation}
One can now form the average with respect to $\hat\rho_0$, to obtain
\begin{eqnarray}
\langle\hat{A}(T)\rangle
& = &  \langle\hat{A}\rangle_0+i\int_{-T}^0
dt'\langle[\hat{H},\hat{A}(t')]\rangle_0\nonumber \\
& = & \langle\hat{A}\rangle_0+i\int_{-T}^0
dt'\langle[\hat{V},\hat{A}(t')]\rangle_0.
\label{v:a}
\end{eqnarray}
For the existence of a well-defined limit  $\langle
\hat{A}(\infty)\rangle$, one must show that~\cite{taylor}
\begin{equation}\label{convergence}
\int_{-\infty}^0 dt\langle[\hat{V},\hat{A}(t)]\rangle_0
<+\infty\;\;.
\end{equation}
To this end one argues
that as long as  $\hat{V}$ and $\hat{A}$ are operators local to
the quantum dot,\cite{comment1}
the time-evolution of $\hat{A}(t)$ will decay
as electrons travel away and the integral is
finite. 

To make the argument  concrete, we consider as example the usual on-site
Coulomb interaction $\hat{V}=Un_{d\uparrow}n_{d\downarrow}$ and measure
the current through the dot, $\hat{A}=\hat{I}$. The occupation number
operator can be expressed in terms of $\psi^\dagger_{\alpha k\sigma}$ as
\begin{equation}
\hat{n}_{d\sigma} =\sum_{kk',\alpha\alpha'}\frac{t_\alpha t_{\alpha'}}{\Omega}
g^*_d(\epsilon_k)g_d(\epsilon_k')\psi^\dagger_{\alpha k\sigma}\psi_{\alpha'
k'\sigma}.
\end{equation}
With the requirement that the current through the $L/R$ leads, $I_{L/R}$, is the same, the current 
operator $\hat{I}$ can be symmetrized as
$\hat{I}=
(t_R^2 \hat{I}_L+t_L^2 \hat{I}_R)/(t_L^2+t_R^2)$ and
\begin{eqnarray}
\langle\hat{I}\rangle& =& 
\frac{-it_Lt_R}{\sqrt\Omega(t_L^2+t_R^2)}\nonumber\\
& & \sum_{k\sigma}
[\langle d^\dagger_\sigma (t_Rc_{Lk\sigma}-t_Lc_{Rk\sigma})\rangle-\text{h.c.}] \\
&=&\frac{t_Lt_R}{t_L^2+t_R^2}\frac{i}{\Omega}\sum_{kk'}(g_d^*(k)-g_d(k'))
\nonumber \\
& &\times \left[
t_Lt_R\langle\psi^\dagger_{Lk}\psi_{Lk'}-\psi^\dagger_{Rk}\psi_{Rk'}\rangle\right.
\nonumber \\
& &\left.-(t_L^2-t_R^2)\langle\psi^\dagger_{Lk}\psi_{Rk'}+\psi^\dagger_{Rk}\psi_{Lk'}\rangle
\right].
\label{currentop}
\end{eqnarray}
We evaluate Eq.~(\ref{v:a}) using Wick's theorem. Due to the
commutator inside the expectation value, only connected contractions
between any $\hat{V}$ and $\hat{I}(t)$ will contribute.
Therefore any non-vanishing Wick's contractions must have an
even number of contractions connecting $\hat{V}$ and $\hat{I}(t)$ and
contain a factor $\langle
\psi_{\alpha k \sigma}(0)\psi^\dagger_{\alpha k \sigma}(t)\rangle_0$ or
$\langle
\psi^\dagger_{\alpha k \sigma}(0)\psi_{\alpha k \sigma}(t)\rangle_0$.
More specifically, the first order perturbation involves factors like
\begin{eqnarray}
&& \langle[\hat{V},\hat{I}_0(t)]\rangle_0 
\propto \frac{1}{\Omega^2}\sum_{kk'}(g_d^*(k)-g_d(k'))
g_d(k)g_d^*(k')
\nonumber \\
&& 
\times (f_{Lk}-f_{Lk'})e^{-i(\epsilon_k-\epsilon_k')t}+\cdots.
\end{eqnarray}
Summation over the continuum variables $k,k'$ leads to terms of the form
\begin{eqnarray}
\langle d^\dagger(t)d(0)\rangle&=&\frac{1}{\Omega}\sum_k g_d(k)f_\alpha(k)e^{-i\epsilon_k t}
\nonumber \\
&\leq&
\frac{1}{\Omega}\sum_k g_d(k)e^{-i\epsilon_k t}\nonumber\\
&\propto&
e^{-\Gamma |t|}.
\label{eq:decay}
\end{eqnarray}
{Note that the
inequality holds both for
equilibrium and nonequilibrium.}
Therefore, the following expression 
\begin{eqnarray}
\langle[\hat{V}(s_k),[\cdots,[\hat{V}(s_1),\hat{I}_0(t)]\cdots]\rangle_0
\nonumber \\
\propto e^{-\Gamma\cdot{\rm min}\{|s_1-t|,\cdots,|s_k-t|\}}
\end{eqnarray}
holds to any order of the
perturbative expansion in $V$, 
and the integral over $t$, Eq.~(\ref{convergence}), becomes convergent.
This shows
that the steady-state limit of the nonequilibrium is well-defined
regardless of the adiabatic rate $\eta$.

However, it should be emphasized that, although the convergence factor
$e^{\eta t}$ is not necessary for a time-dependent theory, such
adiabatic factor should be treated carefully in a time-independent
theory, like the steady-state nonequilibrium. Such
situation arises in particular when we perform a
Fourier transformation to represent a steady-state quantity in a
spectral representation with sinusoidal basis.
For instance, let us express a steady-state quantity $A$ as an integral
over a time-dependent function $F(t)$,
\begin{equation}
A=\int_{-\infty}^0 F(t)dt,
\end{equation}
where the integral is absolutely convergent without
any adiabatic factor $e^{\eta t}$. We write $F(t)$ in a spectral
representation as
\begin{equation}
F(t) =
\int_{-\infty}^\infty\frac{d\omega}{2\pi}\tilde{F}(\omega)e^{-i\omega t}\;,
\end{equation}
with the Fourier component $\tilde{F}(\omega)$, and the quantity
$A$ becomes
\begin{equation}
A=\int_{-\infty}^0dt\left[\int_{-\infty}^\infty\frac{d\omega}{2\pi}
\tilde{F}(\omega)e^{-i\omega t}\right]\;\;.
\end{equation}
If we now want to express $A$ via a spectral representation, we need to
change the order of integrals. However, $e^{-i\omega t}$ is an
oscillatory function and we have to insert a regularization factor
$e^{\eta t}$ to unambiguously allow the integral exchange. Then
\begin{eqnarray}
A&=&\int_{-\infty}^\infty\frac{d\omega}{2\pi}\tilde{F}(\omega)\left[
\int_{-\infty}^0dt
e^{-i(\omega+i\eta) t}\right]\nonumber \\
&=&\int_{-\infty}^\infty\frac{d\omega}{2\pi}
\frac{i\tilde{F}(\omega)}{\omega+i\eta}\;\;,
\end{eqnarray}
where the limit $\eta\to0$ has to be taken \emph{after} the integral has been evaluated.

Thus, 
the regularization factor
$i\eta$
appears explicitly in the theory. A possible way to avoid it is 
to use an imaginary-time formulation, which is built
on a finite contour cut off by a finite temperature and therefore does not need such a
regularization factor.
It is thus
one of our goals 
to clarify under what conditions a regularization is
not necessary and justify the use of an imaginary-time theory.

\subsection{Conventional analytic continuation}
\label{sec:conventional}

\begin{figure}
\rotatebox{0}{\resizebox{3.0in}{!}{\includegraphics{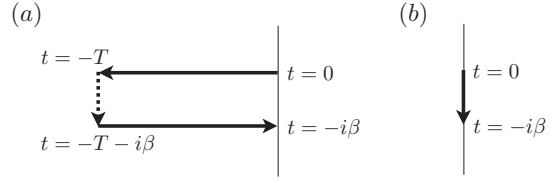}}}
\caption{(a) Keldysh contour for real-time diagrammatics. If the
time-evolution along the dashed line does not contribute an extra
factor, the whole contour can be deformed to one along the
imaginary-time from $t=-i\beta$ to $t=0$ as shown in (b).
}
\label{realimaginary}
\end{figure}

In this subsection, we discuss conventional arguments of the analytic
continuation of a real-time theory to an imaginary-time theory. We furthermore 
illustrate why such deformation of time-contour fails for a steady-state
nonequilibrium, closely following the argument by Doyon and
Andrei~\cite{doyon}. 

In equilibrium, the thermal average of
an observable $\hat{A}$ is given as
\begin{equation}
\langle \hat{A}\rangle = \lim_{T\to\infty}\frac{{\rm
Tr}S(0,-T)\hat\rho_0S(-T,0)\hat{A}}{
{\rm Tr}S(0,-T)\hat\rho_0 S(-T,0)},
\label{eq:A}
\end{equation}
with the time-evolution operator $S(t_1,t_2)=e^{-itH(t_1-t_2)}$ with the
full Hamiltonian $\hat{H}$ and the non-interacting density matrix
$\hat\rho_0=e^{-\beta\hat{H}_0}$. We consider that the limit $T\to\infty$
exists as discussed in the previous section. In the interaction picture
with $\hat{V}_I(t)=e^{it\hat{H}_0}\hat{V}e^{-it\hat{H}_0}$, the above
relation can be rewritten as
\begin{equation}
\langle \hat{A}\rangle = \lim_{T\to\infty}\frac{{\rm 
Tr}S_I(0,-T)\hat\rho_0S_I(-T,0)\hat{A}}{
{\rm Tr}S_I(0,-T)\hat\rho_0 S_I(-T,0)},
\label{eq:exp}
\end{equation}
with
\begin{equation}
S_I(t_1,t_2)={\cal T}\exp\left[-i\int_{t_2}^{t_1}ds \hat{V}_I(s)\right],
\end{equation}
with the time-ordering operator ${\cal T}$ defined as the time moving in
the direction $t_2\to t_1$.
Using the relation,
\begin{equation}
S_I(b,a)=e^{-icH_0}S_I(b+c,a+c)e^{icH_0},
\end{equation}
one can write
\begin{equation}
S_I(0,-T)\hat\rho_0=\hat\rho_0 S_I(-i\beta,-i\beta-T),
\end{equation}
in the similar manner as Ref.~\onlinecite{doyon}. Then $\langle
\hat{A}\rangle$ is written as
\begin{eqnarray}
\langle \hat{A}\rangle &\!\!=\!\!& \lim_{T\to\infty}\frac{{\rm 
Tr}\hat\rho_0S_I(-i\beta,-i\beta-T)S_I(-T,0)\hat{A}}{
{\rm Tr}\hat\rho_0 S_I(-i\beta,-i\beta-T)S_I(-T,0)}\nonumber \\
&\!\!=\!\!& \lim_{T\to\infty}\frac{\langle
S_I(-i\beta,-i\beta-T)S_I(-T,0)\hat{A}\rangle_0}{
\langle S_I(-i\beta,-i\beta-T)S_I(-T,0)\rangle_0}\nonumber\\
&&
\end{eqnarray}
If we can insert the factor $S_I(-i\beta-T,-T)$ [denoted as dashed line
in Fig.~\ref{realimaginary}(a)] between $S_I(-i\beta,-i\beta-T)$ and $S_I(-T,0)$,
one can close the time-contour and analytically continue to the contour
along the imaginary-time $(0,-i\beta)$ [Fig.~\ref{realimaginary}(b)]. 

Using the Wick's
theorem and the linked-cluster theorem, the terms contributing to $\langle
\hat{A}\rangle$ are of the type
\begin{equation}
\langle V_I(s_1)V_I(s_2)\cdots V_I(s_n)\hat{A}(0)\rangle_{0,{\rm connected}}\;,
\end{equation}
where the time $s=0$ and the interaction times $\{s_1,\cdots,s_n\}$ 
are all interconnected by Wick's contractions. When the
interaction $\hat{V}$ and the observable $\hat{A}$ are
operators local to the QD, one can use the relation
Eq.~(\ref{eq:decay}).
{We consider a case that one of $s_k$ in $\langle
V_I(s_1)\cdots V_I(s_n)\hat{A}\rangle_{0,{\rm con}}$ belongs in the
interval $[-T,-i\beta-T]$. In its connected Wick's contractions 
the operators in $\hat{A}$ may be eventually linked to $s_k$
via a forward sequence $\{s'_0=0,\cdots,s'_{p-1},s'_p=s_k\}$ and a
backward sequence $\{s''_0=s_k,\cdots,s''_{q-1},s''_q=0\}$. For the forward sequence
$\{s'_0=0,\cdots,s'_{p-1}\}$ with the times on the real-axis, we can use Eq.~(\ref{eq:decay}),
\begin{equation}
e^{-\Gamma\sum_{n=1}^{p-1}|
s'_n-s'_{n-1}|}
\sim e^{-\Gamma{\rm max}\{|s'_1|,\cdots,|s'_{p-1}|\}}.
\end{equation}
Similar expression holds for the backward sequence.
For the last term involving $s_k\in[-T,-i\beta-T]$, we have a contraction of
$\langle d(s''_1)d^\dagger(s_k)\rangle\langle d(s_k)d^\dagger(s'_{p-1})\rangle
$. For $-\beta<{\rm Im}(s_k)<0$, the two factors remain finite and give
a contribution proportional to $e^{-\Gamma(|
T+s'_{p-1}|+|T+s''_1|)}$.
}
Therefore, when one of the interaction events occurs on the
contour $[-T,-i\beta-T]$, the corresponding  term becomes exponentially small. 
{
When traced with local operator $\hat{A}$, the factorization
property~\cite{doyon} holds
\begin{equation}
S_I(-i\beta,-i\beta-T)S_I(-T,0)\to S_I(-i\beta,0).
\end{equation}}
This
shows that the Wick rotation between real-time and
imaginary-time theory is valid in equilibrium and
\begin{equation}
\langle \hat{A}\rangle = \frac{\langle
S_I(-i\beta,0)\hat{A}\rangle_0}{
\langle S_I(-i\beta,0)\rangle_0}\;\;.
\end{equation}

Next we ask whether the same argument can be extended to the
steady-state nonequilibrium with the initial density matrix at time
$t=-T$ given by $\hat\rho_0=e^{-\beta(\hat{H}_0-\Phi\hat{Y}_0)}$. In order to
move $\hat\rho_0$ 
in Eq.~(\ref{eq:A}) to the leftmost position in the trace, we write $\hat{H}=\hat{H}^\Phi
+\hat{V}^\Phi$ with $\hat{H}^\Phi_0=\hat{H}_0-\Phi\hat{Y}_0$ and 
$\hat{V}^\Phi=\hat{V}+\Phi\hat{Y}_0$. Defining
$V^\Phi_I(t)=e^{it\hat{H}^\Phi_0}\hat{V}^\Phi e^{-it\hat{H}^\Phi_0}$, we
can utilize the same argument as before to write
\begin{equation}
\langle \hat{A}\rangle =
\lim_{T\to\infty}\frac{\langle
S^\Phi_I(-i\beta,-i\beta-T)S^\Phi_I(-T,0)\hat{A}\rangle_0 }{
\langle S^\Phi_I(-i\beta,-i\beta-T)S^\Phi_I(-T,0)\rangle_0}\;\;.
\end{equation}
However, unlike in equilibrium, we cannot use Eq.~(\ref{eq:decay}) for
a contraction containing $V^\Phi_I(s)$ since $\hat{V}^\Phi=\hat{V}
+\Phi\hat{Y}_0$ contains spatially extended operators $c^\dagger_{\alpha k\sigma}
c_{\alpha' k'\sigma'}$ with contributions well away from the QD. Furthermore, $V^\Phi_I(s)
=e^{is\hat{H}^\Phi_0}\hat{V}e^{-is\hat{H}^\Phi_0}+\Phi\hat{Y}_0$ with a
constant of motion $\hat{Y}_0$ with respect to $\hat{H}^\Phi_0$, and
$V^\Phi_I(s)$ would never lead to an exponential decay
for the interactions occurring on the dashed contour in
FIG.~\ref{realimaginary}(a). This shows that a straightforward analytic
continuation of the nonequilibrium Keldysh contour to an imaginary-time one is not possible.

\subsection{Matsubara voltage}

Recently, one of the authors and Heary~\cite{prl07} proposed that, by introducing a
Matsubara term to the source-drain voltage, one can extend the
equilibrium formalism such that the perturbation expansion of the
imaginary-time Green function can be mapped to the Keldysh real-time
theory. The unperturbed Hamiltonian is written as
\begin{equation}
\hat{K}_0(i\varphi_m)=\hat{H}_0+(i\varphi_m-\Phi)\hat{Y}_0,
\end{equation}
with the \textit{Matsubara voltage} $\varphi_m=4\pi m/\beta$ with
integer $m$. We take the many-body interaction $\hat{V}$ as
perturbation.

The non-interacting Hamiltonian
appears in the perturbative expansion in two ways: first in the thermal
factors $e^{-\beta\hat{K}_0}$, and second in the time-evolution
$e^{-\tau\hat{K}_0}$ for the imaginary-time variable $\tau\in[0,\beta)$.
The main trick of this formalism is that in the thermal factor
$i\varphi_m$-dependence drops out as follows. Since
$[\hat{H}_0,\hat{Y}_0]=0$,
$e^{-\beta\hat{K}_0}=e^{-\beta(\hat{H}_0-\Phi\hat{Y}_0)}
e^{-i\varphi_m\beta\hat{Y}_0}$. Since, with respect to the
non-interacting scattering state basis, $\hat{Y}_0$ is diagonal and has
(half)-integer eigenvalues, $e^{-i\varphi_m\beta\hat{Y}_0}=1$,
and 
we have the important identity
\begin{equation}
e^{-\beta\hat{K}_0(i\varphi_m)}=e^{-\beta(\hat{H}_0-\Phi\hat{Y}_0)}=\hat\rho_0\;\;.
\label{identity}
\end{equation}
Therefore, the equivalence of the imaginary-time and real-time formalism
crucially rests on how the double analytic continuation
$i\varphi_m-\Phi\to 0$ and $\tau\to it$ is performed. Since the
$i\varphi_m$-dependence in the thermal factor completely drops out, the
analytic continuation only concerns the time-evolution.
{For
$\tau\in[0,\beta)$, $e^{-i\varphi_m\tau\hat{Y}_0}\neq 1$ and
$i\varphi_m$-dependence does not drop out.} Thus, one could argue
that as $i\varphi_m-\Phi\to 0$ and $\tau\to it$ are taken in that order,
\begin{equation}
e^{-\tau[\hat{H}_0+(i\varphi_m-\Phi)\hat{Y}_0]}
\to
e^{-\tau\hat{H}_0}\to e^{-it\hat{H}_0}.
\end{equation}
However, as we will point out in detail later, integrals over interaction times may create energy
denominators of the type $(K_n-K_m)^{-1}$ in the perturbation
expansions, with $K_n$ being the $n$-th
eigenvalue of $\hat{K}_0$. In such cases, the 
details of the path in the complex plane, along which the 
analytic continuation  $\epsilon_\varphi\equiv i\varphi_m-\Phi\to \pm
i0^+$ is taken, become relevant. On the other hand, in the real-time theory, the
convergence factor $i\eta$ in the energy denominators determines what
poles should be chosen.

\section{Perturbation expansion}\label{sec:III}

\subsection{Real-time expansion}
\label{sec:real}

\begin{figure*}
\myfonts\captionline
\begin{center}
\rotatebox{0}{\resizebox{5.1in}{!}{\includegraphics{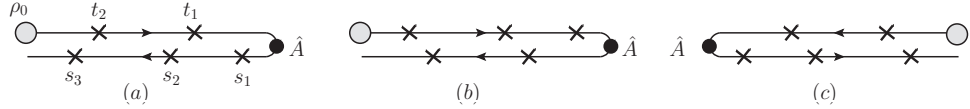}}}
\caption{\small (a) Keldysh contour in forward direction. Crosses mark
interaction points $\hat{V}$ and the dot an observable $\hat{A}$. (b)
Reversed series of scattering points. (c) Backward Keldysh contour
with scattering events equivalent to (a) if $\hat{A}$ is written in
terms of QD operators.}
\label{fig:scatter}
\end{center}
\end{figure*}

In this section, we investigate under what conditions the role of the
regularization factor $\eta$ of the time-independent real-time theory
becomes unimportant.
We assume that a perturbation expansion of  Eq.~(\ref{eq:exp}) exists.
To illustrate the mathematical structure we choose the fifth-order
contribution (as shown in FIG.~\ref{fig:scatter}) and introduce a spectral
representation with respect to the non-interacting scattering state
basis. For the particular time-ordering considered in
FIG.~\ref{fig:scatter}(a), the expression reads
\begin{eqnarray}
S_a&=&(-i)^5{\rm Tr}\left[
\int_0^{-\infty}ds_3\int_0^{s_3}ds_2\int_0^{s_2}ds_1\right.\nonumber \\
&&
\hat{V}_I(s_3)\hat{V}_I(s_2)\hat{V}_I(s_1) \hat{A} \nonumber \\
&&\left.\int_{-\infty}^0dt_2\int_{t_2}^0dt_1\hat{V}_I(t_1)\hat{V}_I(t_2)
\hat\rho_0\right].
\end{eqnarray}
Here we use the notation for intermediate times such that $t_i$ are for
the forward contour ($-\infty\to 0$, upper time contour) and $s_i$ for
the backward (lower) contour.
We redefine the
time as $t_1'=t_1$, $t_2'=t_2-t_1$, $t_i'=t_i-t_{i-1}$ etc., and the
upper part of the Keldysh contour becomes 
\begin{eqnarray}
&&\int_{-\infty}^0dt_2\int_{t_2}^0dt_1\hat{V}_I(t_1)\hat{V}_I(t_2)\nonumber\\
&=&
\int_{-\infty}^0dt_2'\int_{-\infty}^0dt_1'
\hat{V}_I(t_1')\hat{V}_I(t_1'+t_2') \\
&=&
\int_{-\infty}^0dt_2'\int_{-\infty}^0dt_1'
e^{iH_0t_1'}\hat{V}e^{iH_0t_2'}\hat{V}.
\nonumber
\end{eqnarray}
For a spectral representation with respect to energy eigenstates, we
introduce the convergence factor $e^{\eta(t_1'+t_2')}$ for the reasons
discussed in section IA. Then with respect to the non-interacting Fock
basis $|n\rangle$ and $|p\rangle$,
\begin{widetext}
\begin{equation}
(-i)^2\langle
p|\int_{-\infty}^0dt_2\int_{t_2}^0dt_1\hat{V}_I(t_1)\hat{V}_I(t_2)|n\rangle
=\sum_{q} \frac{V_{pq}V_{qn}}{(E_n-E_p+i\eta)(E_n-E_q+i\eta)}\;.
\end{equation}
One can do the same for the lower part of the Keldysh contour,
\begin{eqnarray}
(-i)^3\langle
n|\int_0^{-\infty}ds_3\int_0^{s_3}ds_2\int_0^{s_2}ds_1
\hat{V}_I(s_3)\hat{V}_I(s_2)\hat{V}_I(s_1)|l\rangle 
&=&\nonumber\\
&&\hspace{-4cm}\sum_{mk} \frac{V_{nm}V_{mk}V_{kl}}{(E_n-E_m-i\eta)(E_n-E_k-i\eta)(
E_n-E_l-i\eta)}\;.
\end{eqnarray}
Therefore the above expression $S_a$ can be written as
\begin{equation}
S_a=\sum_{nmklpq}
\frac{V_{nm}V_{mk}V_{kl}}{
(E_n-E_m-i\eta)
(E_n-E_k-i\eta)
(E_n-E_l-i\eta)}A_{lp}
\frac{V_{pq}V_{qn}}{
(E_n-E_p+i\eta)
(E_n-E_q+i\eta)}
\rho_n.
\label{eq:sa}
\end{equation}
Note that all energy denominators consist of one energy \textit{anchored} at
$|n\rangle$ where $\hat\rho_0$ acts at $t=-\infty$ and the other energy of intermediate
states $|m,k,l,p,q\rangle$. For the forward contour, the state
$|n\rangle$ contributes the energy $E_n+i\eta$ in the energy
denominator, and $E_n-i\eta$ for the backward contour.

We now consider a counter-time-ordering as depicted in
FIG.~\ref{fig:scatter}(b) where the
number of scattering events on the lower and upper branches are swapped.
After an explicit calculation by applying the same rules as before, one
gets
\begin{equation}
S_b=\sum_{nmklpq}
\frac{V_{nq}V_{qp}}{
(E_n-E_q-i\eta)
(E_n-E_p-i\eta)}A_{pl}
\frac{V_{lk}V_{km}V_{mn}}{
(E_n-E_l+i\eta)
(E_n-E_k+i\eta)
(E_n-E_m+i\eta)}
\rho_n.
\label{eq:sb}
\end{equation}
\end{widetext}

Starting with the state $|n\rangle$, the numerator $V_{nq}V_{qp}A_{pl}V_{lk}V_{km}V_{mn}\rho_n$ in
Eq.~(\ref{eq:sb}) represents the
reversed process of $\rho_n V_{nm}V_{mk}V_{kl}A_{lp}V_{pq}V_{qn}$ in
Eq.~(\ref{eq:sa}). The factor $\rho_n V_{nm}V_{mk}V_{kl}A_{lp}V_{pq}V_{qn}$ is
understood as the amplitude of the following process
\begin{eqnarray}
S_a: && 
|n\rangle \xrightarrow{\hat{V}}
|q\rangle \xrightarrow{\hat{V}}
|p\rangle \xrightarrow{\hat{A}}\nonumber\\
&&\;\;\;
|l\rangle \xrightarrow{\hat{V}}
|k\rangle \xrightarrow{\hat{V}}
|m\rangle \xrightarrow{\hat{V}}
|n\rangle.
\label{process}
\end{eqnarray}
The many-body interaction can be
written in terms of four scattering state operators as $\hat{V}=\sum
v_{1234}\psi^\dagger_1\psi^\dagger_2\psi_3\psi_4$. With the on-site
Coulomb interaction,
\begin{equation}
\hat{V}=U\sum_{\{\alpha,k\}}
t_1t_2t_3t_4
g^*_1g_2g^*_3g_4
\psi^\dagger_{1 \uparrow}
\psi_{2 \uparrow}
\psi^\dagger_{3 \downarrow}
\psi_{4 \downarrow},
\end{equation}
where the shorthand notations $t_i=t_{\alpha_i}/\sqrt\Omega$,
$g_i=g_d(k_i)$ and
$\psi^\dagger_{i\sigma}=\psi^\dagger_{\alpha_i k_i \sigma}$ have been
used. Note that any creation of a particle $\psi^\dagger_i$ is associated with
the factor $t_ig^*_i$, and the annihilation $\psi_j$ with $t_jg_j$.
For the observable
$\hat{A}$ we consider a one-body operator
$\hat{A}=\sum a_{12}\psi^\dagger_1\psi_2$ for simplicity.
The operator $\hat{V}$ creates up to two particle-hole pairs
of type $\psi$, and for a non-zero matrix element $\langle n|V|m\rangle$, 
$|n\rangle$ and $|m\rangle$ differ only
by up to one particle-hole pair per spin channel. Thus, in the above process
Eq.~(\ref{process}), which starts and
ends with $|n\rangle$, the product of creation operators
$\psi^\dagger_{\alpha k\sigma}$ must match the that of
annihilation operators $\psi_{\alpha k\sigma}$.
Therefore, the matrix element for the process Eq.~(\ref{process}) 
must be of the form
\begin{equation}
S_a:
|t_1g_1|^2 |t_2g_2|^2\cdots t_ig_i a_{ij} t_jg_j^*.
\end{equation}

Similarly, the process for $S_b$-term
\begin{eqnarray}
S_b&:&
|n\rangle \xrightarrow{\hat{V}}
|m\rangle \xrightarrow{\hat{V}}
|k\rangle \xrightarrow{\hat{V}}\nonumber\\
&&\;\;\;|l\rangle \xrightarrow{\hat{A}}
|p\rangle \xrightarrow{\hat{V}}
|q\rangle \xrightarrow{\hat{V}}
|n\rangle
\end{eqnarray}
must contain the same set of $\{\psi^\dagger,\psi\}$ with the same
states, only in the reversed order. The matrix
element for the process then becomes
\begin{equation}
S_b:
|t_1g_1|^2 |t_2g_2|^2\cdots
t_jg_ja_{ji}t_ig^*_i.
\end{equation}
If the operator $\hat{A}$ satisfies the following property
\begin{equation}
g_d(k_i)a_{ij}[g_d(k_j)]^*=g_d(k_j)a_{ji}[g_d(k_i)]^*,
\label{Acondition}
\end{equation}
the matrix elements for counter-contours (a) and (b) match, \textit{i.e.}
\begin{equation}
V_{nm}V_{mk}V_{kl}A_{lp}V_{pq}V_{qn}
=V_{nq}V_{qp}A_{pl}V_{lk}V_{km}V_{mn}.
\label{eq:reverse}
\end{equation}
With this condition, $S_a(\eta)=S_b(-\eta)$, and 
$S_a+S_b$, inside the expression for $\langle \hat{A}\rangle$,  is independent of the sign of
$\eta$ and has a well-defined limit of $\eta\to\pm 0$. {
The above argument can be repeated for
any order of the perturbation expansion, i.e.\ the use of a spectral representation is permitted and
the result independent of the convergence factor $\eta$ provided} that 
the contour has
itself as the counter-contour, $S_a(\eta)=S_a(-\eta)$.

Which of the physically interesting operators do satisfy the above condition Eq.~(\ref{Acondition}) 
respectively (\ref{eq:reverse})? It is easy to see that it is
true for any operator
$\hat{A}$ which is a simple function of $n_{d\sigma}=d^\dagger_\sigma d_\sigma$.
A general two-body operator
$$
\hat{A}=\sum_{1234}a_{1234}\psi^\dagger_1\psi^\dagger_2\psi_3\psi_4
$$
also falls into this class if it satisfies
\begin{eqnarray}
g_d(k_i)g_d(k_j)a_{ijnm}[g_d(k_n)g_d(k_m)]^*\nonumber \\
=g_d(k_n)g_d(k_m)a_{nmij}[g_d(k_i)g_d(k_j)]^*.
\end{eqnarray}
Unfortunately, the current operator Eq.~(\ref{currentop})
does not satisfy the 
condition Eq.~(\ref{Acondition}), and a direct analytic continuation is
not available, as we will discuss shortly. Therefore, we have to resort
to the Meir-Wingreen formula,~\cite{wingreen} which relates the current to the spectral function.

We have so far ignored coinciding energy denominators in the
perturbation expansion leading to overlapping
$\delta$-functions.
For the sake of simplicity we consider a second-order contribution from
Eq.~(\ref{eq:exp}). By expanding it into different time-orderings, we obtain
\begin{eqnarray}
\int^T_0 dt_1\int^T_{t_1}dt_2
\hat\rho_0\hat{V}_I(t_2)\hat{V}_I(t_1)\hat{A}\nonumber \\
+\int_T^0 dt_1\int^T_0dt_2
\hat{V}_I(t_1)\hat\rho_0\hat{V}_I(t_2)\hat{A}\nonumber \\
+\int_T^0 dt_1\int_T^{t_1}dt_2
\hat{V}_I(t_1)\hat{V}_I(t_2)\hat\rho_0\hat{A}.
\label{timedep}
\end{eqnarray}
We now introduce the convergence factor $e^{\eta t}$ and take
$T\to\infty$ to obtain the expression
\begin{eqnarray*}
&&\sum_{nml}\left[
\frac{\rho_n}{(E_n-E_m+i\eta)(E_n-E_l+i\eta)}\right. \nonumber \\
&&+\frac{\rho_m}{(E_m-E_n+i\eta)(E_m-E_l-i\eta)} \\
&&\left.+\frac{\rho_l}{(E_l-E_n-i\eta)(E_l-E_m-i\eta)}
\right]V_{nm}V_{ml}A_{ln}.\nonumber
\end{eqnarray*}
which needs precaution when the two energies in the denominators become equal,
because the contribution will be a product of two
$\delta$-functions with the same argument. One must be careful when one
performs the limit $T\to\infty$. To see this let us
go back to the time-dependent description.
By keeping $T$ finite, contributions of the form $\delta(E_n-E_m)^2$
will actually amount to terms proportional to $T^2$ from the integrals.
Combining all three integrals we obtain the coefficient to the
$T^2$-term (\textit{i.e.} $\delta^2$-term) proportional to
\begin{equation}
\begin{array}[b]{ll}\displaystyle
\sum_{nml}& (\rho_n-2\rho_m+\rho_l)V_{nm}V_{ml}A_{ln}\times\\[3mm]
&\displaystyle\delta(E_n-E_m)\delta(E_m-E_l)\end{array}\;\;.
\label{eq:double}
\end{equation}
In equilibrium $\rho_n=\rho_m=\rho_l$ for $E_n=E_m=E_l$  and this term vanishes identically.
The argument can be easily extended to arbitrary orders in the perturbation expansion.

In the case of nonequilibrium the situation is more complex. Here we
discuss in detail what happens to Eq.~(\ref{eq:double}).
We consider the case  $|n\rangle\neq|m\rangle\neq|l\rangle$, while
$E_n=E_m=E_l$. Suppressing the $\delta$-functions, Eq.~(\ref{eq:double})
has the form
$$
e^{-\beta E_n}(e^{\beta\Phi Y_{0n}}
-2e^{\beta\Phi Y_{0m}}
+e^{\beta\Phi Y_{0l}})V_{nm}V_{ml}A_{ln}\;.
$$
In the matrix element $V_{nm}V_{ml}A_{ln}$, the transition
$|n\rangle\to|m\rangle\to|l\rangle\to|n\rangle$ involves a certain series
of particle-hole excitations. For instance, $|n\rangle\to |m\rangle$ is
given by an exchange of two particle-hole pairs, $\psi^\dagger_{\alpha_1k_1\sigma}
\psi_{\alpha_2k_2\sigma}\psi^\dagger_{\alpha_3k_3\sigma'}
\psi_{\alpha_4k_4\sigma'}$
in $\hat{V}$, and similarly for $|m\rangle\to |l\rangle$ and $|l\rangle\to
|n\rangle$. However, since any creation of $\psi^\dagger_{\alpha
k\sigma}$ should be matched by $\psi_{\alpha k\sigma}$ only up to 6
indices are independent. Given a particular set of the 6 indices of
wave-vectors and spins $\{k_1\sigma_1,k_2\sigma_2,\cdots,k_6\sigma_6\}$, 
different permutations of the above 6 pairs of
$\{\psi^\dagger_{k_i\sigma_i},\psi_{k_i\sigma_i}\}$ in
$\hat{V}\hat{V}\hat{A}$ determines the matrix element
$V_{nm}V_{ml}A_{ln}$.
Now, we sum over all possible combinations of reservoir indices
$\{\alpha_1,\cdots,\alpha_6\}$ (while keeping the $k$-indices unchanged) 
for the all twelve $\{\psi^\dagger,\psi\}$ operators. The matrix element
$V_{nm}V_{ml}A_{ln}\propto
\prod_{i=1,6}t^2_{\alpha_i}|g(\epsilon_{k_i})|^2$. Since
the product of $|g(\epsilon_{k_i})|^2$ are invariant, we
collect all possible reservoir weights in $\prod_{i=1,6}t^2_{\alpha_i}e^{\beta\Phi
Y_{0\{n,m,l\}}}$ and each of the three
sums in Eq.~(\ref{eq:double}) become the same, i.e.\ the whole contribution vanishes. A detailed 
discussion of the mathematics can be found in Appendix~\ref{app:double}.

In summary, if the observable $\hat{A}$ satisfies
Eq.~(\ref{Acondition}), the energy integration in the perturbation
expansion can be interpreted as principal-valued, similarly to
equilibrium. In Appendix~\ref{app:fourth}, we provide as an example the
fourth-order contribution to the QD-electrons self-energy and show
explicitly that the above properties are satisfied. 
{ 
Since the structures appearing in higher order 
are of the same type as discussed above, we may actually infer that this property holds in any order
of the perturbation expansion.}

\subsection{Imaginary-time expansion}

Unlike the real-time theory, the imaginary-time description is
formulated on a finite time interval of $[0,\beta)$, and there is no
need for a convergence factor $e^{\eta t}$. Therefore, the energy
integrals appearing in the equilibrium theory are always principal-value
integrals, which we confirmed in the previous
section~\ref{sec:conventional}.

In nonequilibrium, with the imaginary-time effective Hamiltonian
$\hat{K}(i\varphi_m)=\hat{H}_0+\epsilon_\varphi\hat{Y}_0+\hat{V}$
($\epsilon_\varphi=i\varphi_m-\Phi$), the thermal average is defined as
\begin{equation}
\langle{\cal A}\rangle =\frac{{\rm Tr}e^{-\beta\hat{K}}{\cal A}}{
{\rm Tr}e^{-\beta\hat{K}}}.
\end{equation}
The Boltzmann factor can be expanded as
\begin{equation}
e^{-\beta\hat{K}}=e^{-\beta\hat{K}_0}{\cal T}_\tau\exp\left[
-\int_0^\beta d\tau V_I(\tau)\right],
\end{equation}
with $V_I(\tau)=e^{\tau\hat{K}_0}\hat{V}e^{-\tau\hat{K}_0}\hat{V}$ and
${\cal T}_\tau$ denoting the time-ordering operator for $\tau\in[0\to
\beta]$.
We consider a second order expansion to understand its mathematical
structure,
\begin{eqnarray}
& & {\rm Tr}\,e^{-\beta\hat{K}_0}\int_0^\beta d\tau \int_0^\tau
d\tau'V_I(\tau)V_I(\tau')\hat{A} \nonumber \\
& = & \int_0^\beta d\tau \int_0^\tau d\tau'\nonumber \\
& & \sum_{nml}
\rho_n e^{\tau(K_n-K_m)}V_{nm}e^{\tau'(K_m-K_l)}V_{ml}A_{ln}\nonumber \\
&=&\sum_{nml}\left[
\frac{\rho_n}{(K_n-K_m)(K_n-K_l)}\right. \nonumber \\
&&+\frac{\rho_m}{(K_m-K_l)(K_m-K_n)}\nonumber \\
&&\left.+\frac{\rho_l}{(K_l-K_n)(K_l-K_m)}
\right]V_{nm}V_{ml}A_{ln}.
\label{squarebra}
\end{eqnarray} 
This expression has the same mathematical structure as in the real-time
theory.
Even though we considered only one
time-ordering in the imaginary-time theory, the upper and lower integral limits
in $\int_0^\beta d\tau\int_0^\tau d\tau'$ combine to create the same permutation of terms 
as in the real-time theory~\cite{prl07}.

We have seen earlier that, in the real-time theory, energy denominators
can be interpreted as principal-valued since all $\delta$-function
contributions from the energy poles vanish. Therefore,
if we interpret the energy denominators as principal-valued as
$i\varphi_m\to\Phi$
\begin{equation}
\frac{1}{K_n-K_m}\to {\cal P}\left(\frac{1}{E_n-E_m}\right)
\end{equation}
the terms in the imaginary-time theory indeed match those of the
real-time approach.

In section~\ref{sec:double}, we calculate the double occupancy from
continuous-time quantum Monte Carlo method, and numerically verify that the
analytic continuation procedure outlined so far works accurately 
{ 
in all
orders of perturbation theory as well as for the resummed perturbation series}.

\subsection{Single-particle self-energy}\label{sec:SPSE}

The analytic properties discussed so far can be used to examine the
single-particle self-energy for the Anderson impurity model.
The imaginary-time second-order self-energy in the Coulomb interaction $U$ 
can be written as~\cite{prl07}
\begin{equation}
\Sigma^{(2)}(i\omega_n,\epsilon_\varphi)=\sum_\gamma\int
d\epsilon\frac{\sigma_\gamma(\epsilon)}{i\omega_n-\frac{\gamma}{2}\epsilon_\varphi-\epsilon},
\end{equation}
with the spectral function
\begin{widetext}
\begin{equation}
\sigma_\gamma(\omega)=U^2 \left[\prod_{i=1}^3 \int d\epsilon_i
A_0(\epsilon_i)\right]\sum_{\alpha_1+\alpha_2+\alpha_3=\gamma}\,\left[f_1(1-f_2)f_3+
(1-f_1)f_2(1-f_3)\right]\,\delta(\omega-\epsilon_1+\epsilon_2-\epsilon_3)
\end{equation}
\end{widetext}
for the $\gamma$-branch cut ($\gamma=\pm
1,\pm 3$), where
$$
A_0(\epsilon)=\frac{\Gamma/\pi}{(\epsilon-\epsilon_0)^2+\Gamma^2}
$$ 
denotes the non-interacting spectral function 
of the QD level and $f_\alpha=[1+e^{-\beta(\epsilon-\alpha\Phi/2)}]^{-1}$
the Fermi-Dirac factor for the $\alpha$-th reservoir.

Recently, it has been proposed~\cite{han10} that an inclusion of higher-order contributions
will mainly modify the spectral function $\sigma_\gamma(\epsilon)$, leading to 
a $\epsilon_\varphi$ dependence like
\begin{equation}
\Sigma(i\omega_n,\epsilon_\varphi)=\sum_\gamma\int
d\epsilon\frac{\sigma_\gamma(\epsilon,\epsilon_\varphi)}{
i\omega_n-\frac{\gamma}{2}\epsilon_\varphi-\epsilon}\;\;.
\end{equation}
Based on this expression, one can try to fit  $\sigma_\gamma(\epsilon,\epsilon_\varphi)$ 
to the numerical single-particle self-energy generated from quantum
Monte Carlo calculations. {
However, in order } to establish the existence of an analytic
continuation limit 
of the imaginary-time self-energy, one should first
show that the real-time self-energy possesses the analytic property
discussed in the previous section, namely that the energy poles are
principal-valued. The rather lengthy and technical argument is
provided 
in Appendix~\ref{app:fourth} for the fourth-order 
self-energy diagrams. It can be shown explicitly that contributions involving 
{
products of $\delta$-functions with identical argument vanish identically},
resulting in 
the necessary analytic properties discussed in the previous section.

{
Again, investigating the general structures appearing in the perturbation expansion
of the self-energy, we are confident that this property indeed holds in any order and also survives
the resummation of the series. The latter aspect, however, cannot be proven rigorously, but is strongly
supported by the numerical evidence from our Monte-Carlo simulations.}

{ 
In a recent work by Dirks
\textit{et
al.}~\cite{dirks} and a accompanying paper to this work, a general
analytic continuation approach based on the multi-variable complex
function theory and its double analytic continuation of
$(i\omega_n,i\varphi_m)$ have been systematically studied.}

\subsection{Forward and backward steady-state}

We have seen in Section~\ref{sec:real} that we need
Eq.~(\ref{eq:reverse}) for any sequence of matrix elements in order to
establish the equivalence of the real and imaginary-time theory.
In order to close the formal
discussions, let us re-examine the complex
conjugate of the matrix elements in relation to
the forward- and backward-in-time propagation of
scattering state density matrix.

Assume that we propagate a non-interacting density matrix
$\rho_0=\exp[-\beta(H_0-\Phi Y_0)]$ from the initial time $t=-T$ to the
present in the \textit{forward} direction. Then, according to Gell-Mann and
Goldberger~\cite{gellmann}, we obtain
\begin{eqnarray}
\hat{\rho}_{out} & = & \eta\int_0^\infty e^{-i{\cal L}T}\left(
e^{i{\cal L}_0T}\hat{\rho}_0\right)e^{-\eta T}dT \nonumber \\
& = & \eta\int_0^\infty e^{-i{\cal L}T}
\hat{\rho}_0e^{-\eta T}dT\nonumber\\
& = & \frac{\eta}{\eta+i{\cal L}}\hat{\rho}_0\nonumber\\
&=&\hat{\rho}_0+\frac{1}{-{\cal L}+i\eta}{\cal L}_V\hat{\rho}_0\;\;,
\label{eq:outstate}
\end{eqnarray}
with ${\cal L}_V$ the Liouvillian representing the interaction parts not contained in ${\cal L}_0$.
$\hat{\rho}_{out}$ is the fully interacting density matrix at $t=0$ and
$\hat{\rho}_0$ non-interacting density matrix at $t=0$.
The meaning of the above equation is that we unwind
a non-interacting density matrix to a remote time $t=-T$ and re-evolve
it with full interaction to the present time. By taking the average over
the remote time $T$, we filter out transient oscillations.

Alternatively, we can also consider
a backward propagation of density matrix evolving from the
remote future by writing
\begin{eqnarray}
\hat{\rho}_{in} & = & \eta\int_0^\infty e^{i{\cal L}T}\left(
e^{-i{\cal L}_0T}\hat{\rho}_0\right)e^{-\eta T}dT \nonumber \\
& = &
\hat{\rho}_0+\frac{1}{-{\cal L}-i\eta}{\cal L}_V\hat{\rho}_0.
\label{eq:instate}
\end{eqnarray}

If we initially choose $\hat{\rho}_0$ as the density matrix of a
quantum dot system of disconnected dot and
reservoirs, ${\cal L}_V={\cal L}_t+{\cal L}_U$ includes both the hopping to
the leads and the
Coulomb interaction on the dot. We first construct the scattering states with
respect to the hopping, and then with respect to the Coulomb interaction. After  the
first step, the scattering states become~\cite{prb07}
\begin{eqnarray}
\psi^\dagger_{\alpha k\sigma,out}&=&c^\dagger_{\alpha
k\sigma}+\frac{t}{\sqrt\Omega}g_d(k)d^\dagger_\sigma+\cdots
\\
\psi^\dagger_{\alpha k\sigma,in}&=&c^\dagger_{\alpha
k\sigma}+\frac{t}{\sqrt\Omega}g_d(k)^*d^\dagger_\sigma+\cdots,
\end{eqnarray}
and we can construct respective scattering-state density matrices $\hat{\rho}_{0t,out}$ and
$\hat{\rho}_{0t,in}$ with ${\cal L}_V={\cal L}_U$.
The coefficients appearing in front of the dot operators $d^\dagger_\sigma,d_\sigma$
etc.\ for the out and in-scattering states are the complex conjugate of each other.
Therefore, the matrix elements of the interaction $\hat{V}=Un_{d\uparrow}n_{d\downarrow}$,
 written in terms of
$\psi_{\alpha k\sigma,\{out,in\}}$-basis, are complex conjugate to each other,
\textit{i.e.} $V_{nm}=V_{\tilde{n}\tilde{m}}^*$ (with the tilde denoting
the in-scattering basis). 

We can now repeat the arguments from Section~\ref{sec:real} for the backward propagation of the 
density matrix as shown in FIG.~\ref{fig:scatter}(c) and find
\begin{widetext}
$$
S_c=\sum_{nmklpq}
\frac{V_{\tilde{n}\tilde{q}}V_{\tilde{q}\tilde{p}}}{
(E_n-E_q+i\eta)
(E_n-E_p+i\eta)}A_{\tilde{p}\tilde{l}}
\frac{V_{\tilde{l}\tilde{k}}V_{\tilde{k}\tilde{m}}V_{\tilde{m}\tilde{n}}}{
(E_n-E_l-i\eta)
(E_n-E_k-i\eta)
(E_n-E_m-i\eta)}
\rho_{\tilde{n}}.
$$
For observables satisfying $A_{nm}=A_{\tilde{n}\tilde{m}}^*$,
this expression becomes identical to $S_a$ in Eq.~(\ref{eq:sa}).
The same argument holds in any order of the  perturbation expansion, and we have $
{\rm Tr} \hat{A}\hat{\rho}_{out}
={\rm Tr} \hat{A}\hat{\rho}_{in}$
and $\langle \hat{A}\rangle=\frac12(\langle \hat{A}\rangle_{out}
+\langle \hat{A}\rangle_{in})$.
Therefore, from Eqs.~(\ref{eq:outstate}), (\ref{eq:instate}), we have
\begin{eqnarray}
\langle \hat{A}\rangle &= &\langle \hat{A}\rangle_0+\left\langle\hat{A}\frac12\left(
\frac{1}{-{\cal L}+i\eta}+\frac{1}{-{\cal L}-i\eta}
\right){\cal L}_V\hat\rho_0\right\rangle
=
\langle \hat{A}\rangle_0+\left\langle\hat{A}{\cal P}\left(
\frac{1}{-{\cal L}}
\right){\cal L}_V\hat\rho_0\right\rangle\;\;,
\end{eqnarray}
\end{widetext}
i.e., the conditions for replacing the energy
denominators by their principal-values, as discussed in
section~\ref{sec:real}, correspond to a measurement protocol where the observable $\hat{A}$
has the same expectation values with respect to the forward- and backward-propagating
density matrices.

\section{Static Expectation Values}\label{sec:IV}
\subsection{Theoretical background}
\label{subsec:theobackgrnd}

We have shown that steady-state expectation values of certain local
observables $\hat A$  can be obtained from analytical continuation of
expectation values calculated within the imaginary time
Matsubara-voltage formalism.  As long as we know 
{the
analytic structure of these objects}, this can be done easily. However,
for a model with true two-particle interactions, one eventually has to
resort to numerical evaluations, and an analytical continuation in
general requires a more involved computational technique. We therefore
want to provide in the following a representation which allows the use
of standard tools from equilibrium many-body theory.

A numerical method gives $\langle \hat A\rangle(i\varphi_m)$ and let $\langle
\hat A\rangle(z_\varphi)$ be its analytic continuation.
We may write formally
\begin{equation}
\langle \hat A\rangle(z_\varphi)=\langle \hat A\rangle_\text{const}+\chi_A(z_\varphi)
\label{eq:formalstrucObs}
\end{equation}
where the part $\chi_A(z)$ is holomorphic in the upper and lower half plane, with singularities
only on the real axis.
If one can furthermore show that the $z\chi_A(z)$ is non-singular in the limit 
$z_\varphi\to \infty$, one can finally infer that a 
spectral representation with respect to the jump function on the real axis 
exists and hence
\begin{equation}
\langle \hat A\rangle(\imag\varphi_m) =
\langle \hat A\rangle_\text{const} +
\int \frac{\varrho_A(\varphi)}{(\imag\varphi_m - \Phi) - \varphi}
\Dfrtl\varphi
\label{eq:representationStaticObservables}
\end{equation}
Note that the latter property is not necessarily guaranteed and has to be proven individually for
each observable.

Once the validity of the representation
\eqref{eq:representationStaticObservables} is established, one only needs to obtain the
``spectral function'' $\varrho_A(\varphi)$. One evident method to calculate the Matsubara
voltage data $\langle \hat A\rangle(\imag\varphi_m)$ for the
observable
$\hat A$ with respect to the effective system with
non-hermitian
Hamiltonian at Matsubara voltage $\imag\varphi_m$ is via a QMC simulation.\cite{dirks} 
For such data with statistical noise, one then
typically employs a maximum-entropy approach (MaxEnt).\cite{mem} 
The implementation of a MaxEnt 
estimator for the physical expectation value
is rather
straightforward. The 
values for different $\imag\varphi_m$ are truly statistically independent, and only the
variance and correlation between imaginary and real parts of a single
$\imag\varphi_m$ value play a role.
However, one still needs accurate and unbiased measurements of imaginary-voltage data over
a large range of $\varphi_m$.\cite{dirks} This latter requirement
makes the use of a 
continuous-time quantum Monte-Carlo (CT-QMC) algorithm mandatory.
In particular, the necessary estimation of the constant offset $\langle \hat A\rangle_\text{const}$
in Eq.~\eqref{eq:representationStaticObservables} is possible only with CT-QMC,
because at present no direct measurement algorithm for this quantity is available
and one must determine it from the 
tail of $\langle \hat A\rangle(\imag\varphi_m)$ by fitting it to
\begin{equation}\label{eq:offset_fit}
\langle \hat A\rangle(\imag\varphi_m)
\stackrel{m\to\infty}{\to}
\langle \hat A\rangle_\text{const} +
\frac{c_A}{\imag\varphi_m}+\frac{\tilde c_A}{(\imag\varphi_m)^2} + \cdots\;\;.
\end{equation}
In practice, a weighted least-square fit yields
reliable values and error bars for $\langle \hat A\rangle_\text{const}$.
Via Gaussian error propagation it
is then possible to incorporate the uncertainty of $\langle
 \hat A\rangle_\text{const}$ into the covariance
matrix of the quantity
$\langle \hat A\rangle(\imag\varphi_m) - \langle
\hat A\rangle_\text{const}$.\footnote{Note that corrections from error
propagation to off-diagonal covariance matrix elements are neglected. This
may be justified,
because correlations between real and imaginary part of an
effective-equilibrium expectation value are the only nonzero off-diagonal
elements anyway.}

In general, the spectral function $\varrho_A(\varphi)$ needs not to be positive semidefinite, or
show any symmetry relations with respect to $\varphi$.
Since on the other hand the MaxEnt method is only applicable for the inference of positive
definite
functions, a shift function $\varrho_\text{shift}(\varphi)$ of the spectral
function $\varrho_A(\varphi)$
has to be introduced, which makes the to-be-inferred $\varrho_A'(\varphi)
= \varrho_A(\varphi) - \varrho_\text{shift}(\varphi)$
positive. We also employ a symmetry condition
\begin{equation}
\varrho_\text{shift}(\varphi) = \varrho_\text{shift}(-\varphi),
\end{equation}
because this choice is robust with respect to the physical result
\begin{equation}
\begin{split}
\langle \hat A\rangle_\text{phys} & = \frac{1}{2}
\sum_{\alpha=\pm1}\langle \hat A\rangle(\Phi+\alpha\imag \eta) \\
& = 
\langle \hat A\rangle_\text{const} -
\mathcal{P}\!\!\!\!\!\!\int\!\!\Dfrtl{\varphi}\frac{\varrho_A(\varphi)}{\varphi}.
\end{split}
\label{eq:physExpVal}
\end{equation}

In the following we want to prove that the double occupancy or magnetization obey this constraint, i.e.\ 
have a representation,
where $\langle \hat A\rangle_\text{const}$ is a real number, and
$\varrho_A(\varphi)\in\mathbb{R}$ is a real-valued spectral function.

\subsubsection{Double Occupancy}
\label{sec:double}

The double occupancy in Matsubara-voltage representation is defined as
\begin{equation}
D(\imag\varphi_m) := 
\left\langle 
n_{d,\uparrow}n_{d,\downarrow}
\right\rangle_{K(\imag\varphi_m)}\;,
\end{equation}
where the expectation value is taken with respect to the $m$-th effective
equilibrium system.

We will first show that the representation
\eqref{eq:representationStaticObservables} holds for the double
occupancy, i.e.\ that we have indeed
\begin{equation}
D(\imag\varphi_m) = D_{0} + \int \Dfrtl{\varphi}
\frac{\varrho_\text{D}(\varphi)}{\imag\varphi_m - \Phi - \varphi}.
\label{eq:specrepdocc}
\end{equation}
We restrict the discussion to the case of particle-hole symmetry and symmetric coupling to
the leads, $\Gamma_L = \Gamma_R$.
Within the Matsubara-voltage approach, one can -- for fixed $\imag\varphi_m$ --
employ the standard techniques of equilibrium many-body theory
and obtains the standard result\cite{bulla:1998}
\begin{equation}
\begin{split}
D(\imag\varphi_m) =& \langle n_\uparrow \rangle \langle n_\downarrow \rangle
\\
& \hspace{-12mm}+ \frac{1}{\beta U} \sum_{\omega_n} \Sigma(\imag\varphi_m; \imag\omega_n )
G(\imag\varphi_m;\imag\omega_n) \euler{\imag\omega_n \eta}\;.
\end{split}
\label{eq:doubleoccWick}
\end{equation}
Due to particle-hole symmetry, we have $\langle n_\uparrow \rangle \langle
n_\downarrow \rangle= 1/4$. Furthermore, from the discussion in section \ref{sec:SPSE}
we can infer that at least the Green's function decays like
$1/\imag\varphi_m$ 
and hence allows for the existence of a spectral representation
\eqref{eq:specrepdocc}, as long as there is only a single branch cut at $\Im
z_\varphi = 0$.

The real-valuedness of spectral function and constant offset remain to be
shown. The general relation $G(-\imag\varphi_m,-\imag\omega_n)^* =
G(\imag\varphi_m, \imag\omega_n)$ holds for Green's function and self-energy. Inserting
this into Eq.~\eqref{eq:doubleoccWick}, we find
\begin{equation}
D(-\imag\varphi_m)^* = D(\imag\varphi_m).
\label{eq:symmreldocc}
\end{equation}
Consequently, the real part of $D(\imag\varphi_m) - D(-\imag\varphi_m)$
vanishes.
Using the symmetric coupling to the leads, we have an invariance of the
Green's function and self-energy under $(\imag\varphi_m - \Phi)
\leftrightarrow -(\imag\varphi_m - \Phi)$. As a result, $D_0$ is an actual
constant which is obtained for both, upper and lower half plane. 
Due to the symmetry of $\Im D(\imag\varphi_m)$, $D_0$ is real.
By inserting the representation \eqref{eq:specrepdocc} into Eq.~\eqref{eq:symmreldocc} we also see that $\varrho_\text{D}(\omega)$ is real-valued.

For example, let us consider the equilibrium setup, i.e.~$\Phi=0$. At half filling 
and symmetric coupling to the leads, the function 
\begin{eqnarray}
\Re D_{\Phi=0}(\imag \varphi_m) & = & \Re D_{\Phi=0}(-\imag \varphi_m), 
\label{eq:realteildoccprop}
\\
\Im D_{\Phi=0}(\imag \varphi_m) & \equiv & 0.
\label{eq:imagteildoccpropweak}
\end{eqnarray}
This is compatible with a conventional bosonic spectral representation
\begin{equation}
D_{\Phi=0}(\imag\varphi_m) = \int \Dfrtl\varphi \frac{\varrho_D(\varphi)}{\imag
\varphi_m - \varphi} + D_0,
\end{equation}
with an antisymmetric spectral function 
\begin{equation}
\varrho_D(\varphi)=-\varrho_D(-\varphi);\quad \varrho_D(\varphi>0) < 0
\end{equation}
and the offset $D_0 > 0$.
Eq.~\eqref{eq:imagteildoccpropweak} is not evident for asymmetric couplings or off  particle-hole
symmetry, because here $G_0(\imag \varphi_m, \imag\tau)$ is
not real.

\subsubsection{Magnetic Susceptibility}
An observable which is much more sensitive to the Kondo effect is
the magnetization $M:=(\langle n_\uparrow\rangle-\langle
n_\downarrow\rangle)$ 
in the presence 
of a magnetic field $B$ in $z$-direction respectively the magnetic susceptibility $\chi=M/B$
of the quantum dot, because it directly probes
the spin degree of freedom of the dot electrons. In equilibrium, a strong
dependence on the temperature is observed, on the scale of the Kondo
temperature.\cite{hewson}

As for the double occupancy, the validity of a spectral representation
\begin{equation}
M(\imag\varphi_m) = M_{0} + \int \Dfrtl{\varphi}
\frac{\varrho_\text{M}(\varphi)}{\imag\varphi_m - \Phi - \varphi}
\label{eq:specrepmagn}
\end{equation}
can readily be confirmed. Starting from the symmetry
$G(-\imag\varphi_m,-\imag\omega_n)^* =
G(\imag\varphi_m, \imag\omega_n)$, one can again show that
$M(-\imag\varphi_m)^* =
M(\imag\varphi_m)$, and the same arguments apply concerning the interchange
$(\imag\varphi_m -\Phi) \leftrightarrow -(\imag\varphi_m -\Phi)$.

\subsection{Numerical effective-equilibrium data}

Let us now turn to the discussion of actual numerical 
data for
magnetization and double occupancy from the quantum Monte-Carlo
simulations. As the first step, we analyze these data
with respect to the auxiliary  variable $\varphi_m$, 
and want to argue that they have a physical interpretation with respect to the actual 
voltage $\Phi$. 
In particular, the convergence of the numerical procedures described below implies full consistency 
of the Matsubara-voltage formalism with regard to the numerical data. 
\begin{figure*}[htb]
\subfloat[$U=3\Gamma$, $\beta\Gamma=10$]
  {\includegraphics[width=0.49\linewidth]{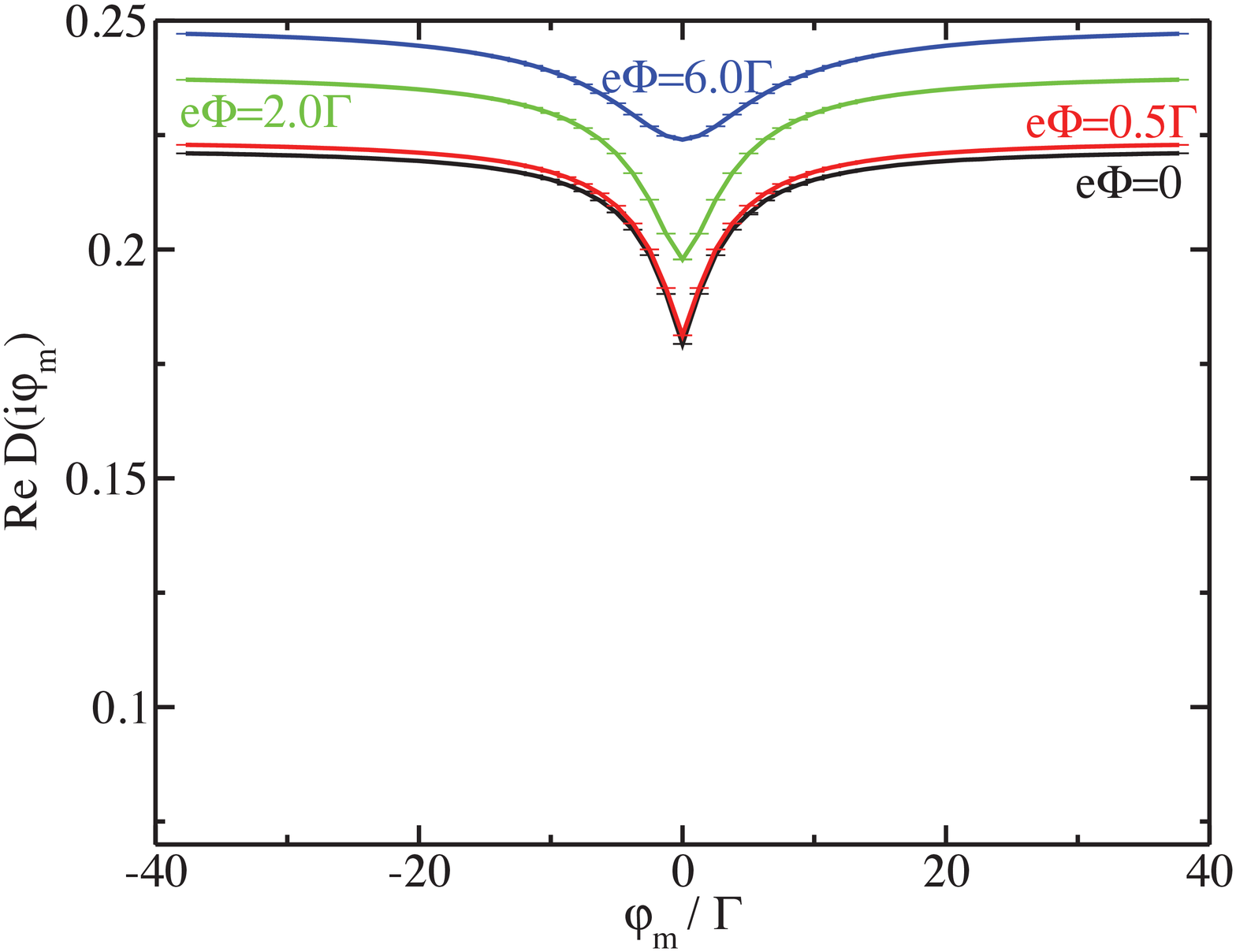}}
\subfloat[$U=5\Gamma$, $\beta\Gamma=20$]
  {\includegraphics[width=0.49\linewidth]{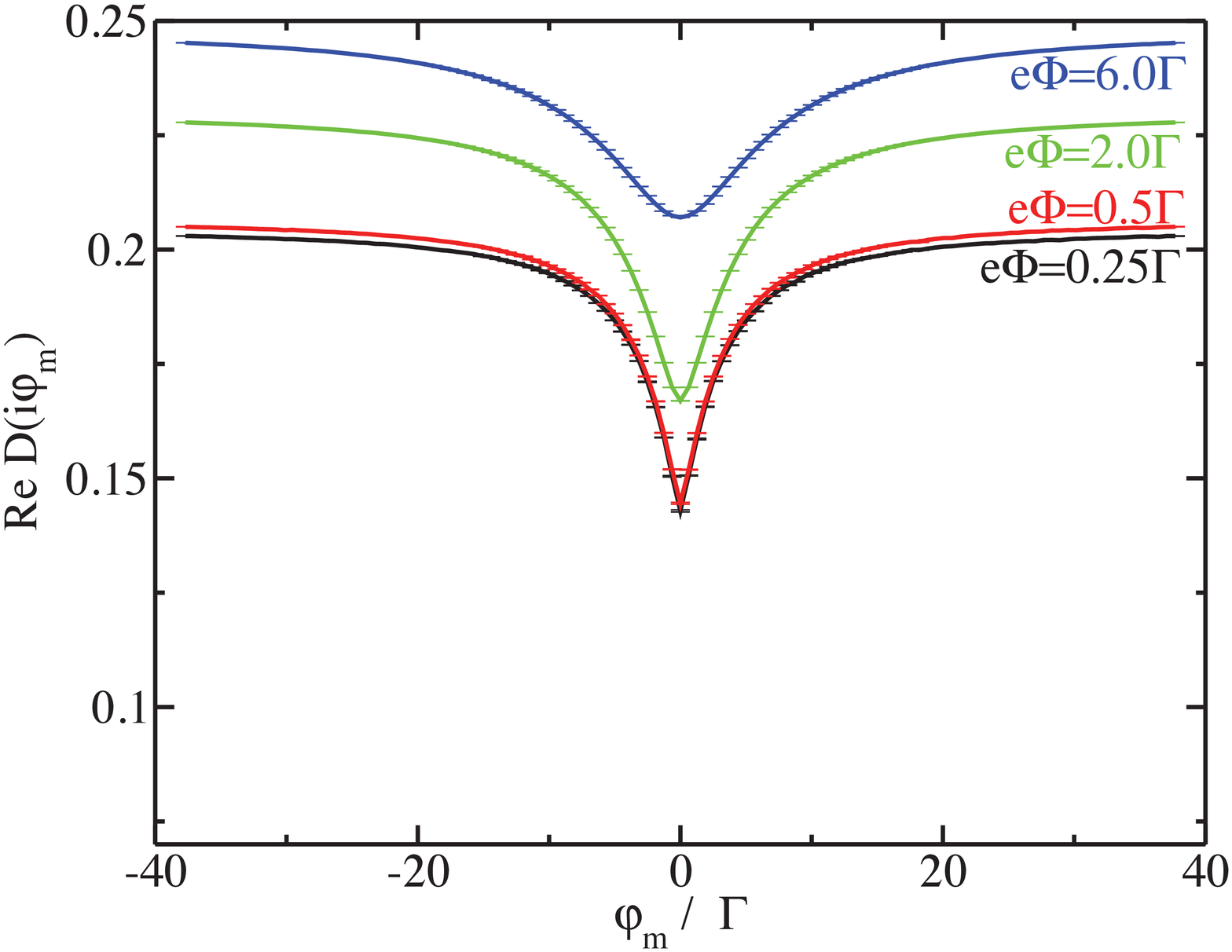}}\\
\subfloat[$U=8\Gamma$, $\beta\Gamma=20$]
  {\includegraphics[width=0.49\linewidth]{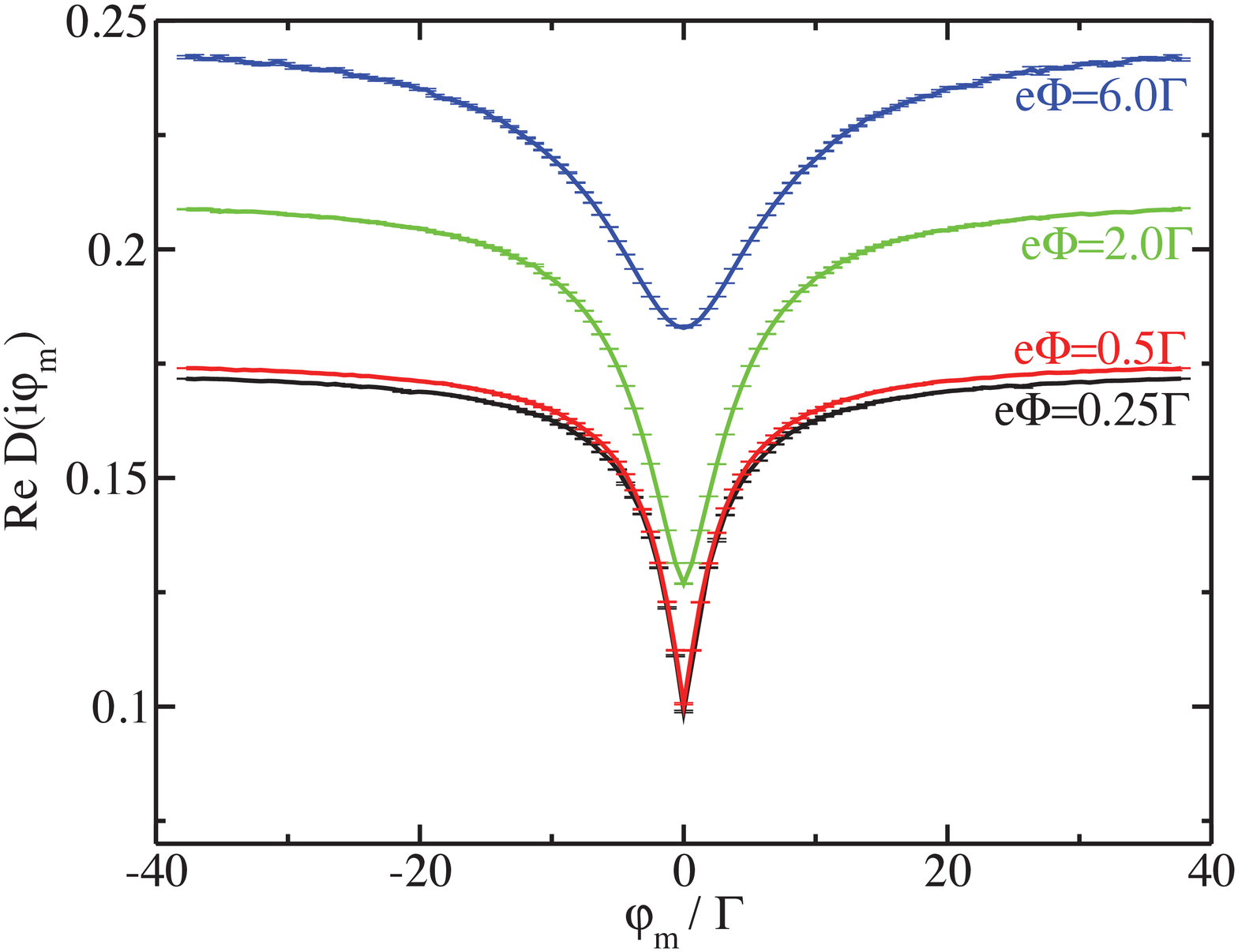}}
\subfloat[$U=10\Gamma$, $\beta\Gamma=20$]
  {\includegraphics[width=0.49\linewidth]{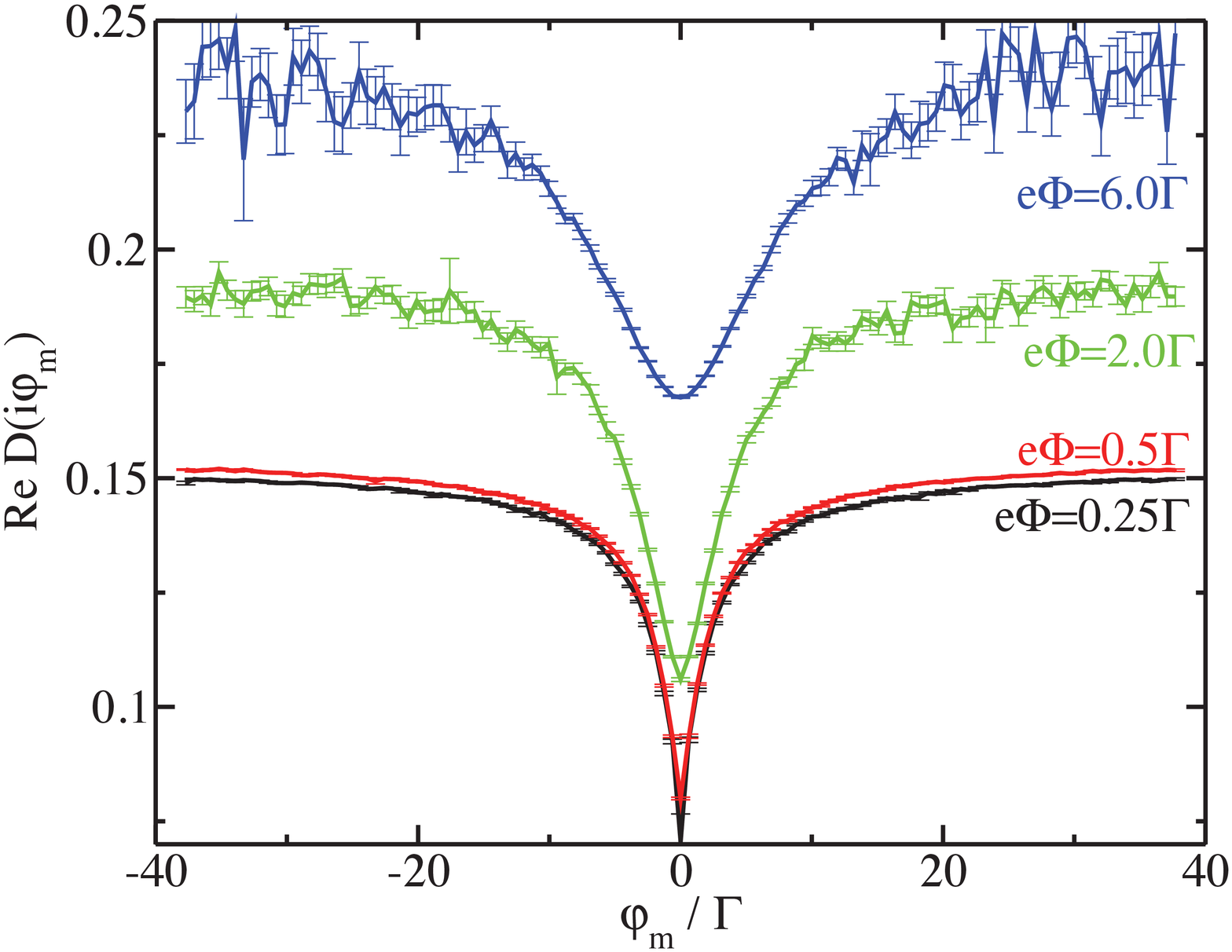}
\label{subfig:rawdatadoubleoccU10}}
\caption{(color online) Real part of the effective-equilibrium double occupancy as a function of the Matsubara
voltage $\varphi_m$ at several values of interaction strength $U$ and bias voltage $\Phi$.}
\label{fig:rawdatadoubleocc}
\end{figure*}

We find that effective-equilibrium data come along with characteristic energy scales
which -- after analytic continuation -- may translate almost directly into energy scales with
respect to the actual source-drain voltage $\Phi$. It is therefore worthwhile to
discuss the dependence of the effective-equilibrium expectation values as a
function of $\varphi_m$ for given physical parameters $\beta$, $U$, and
$\Phi$.

\paragraph{Dependence on $\Phi$.} The first thing to notice is that the dependence of the 
\emph{shape} of the curves
$M(\imag\varphi_m)$ and $D(\imag\varphi_m)$ on $\Phi$ is rather moderate: for the examples
considered, we do not observe any new characteristic energy scales with respect to
the Matsubara voltage $\varphi_m$ emerging or disappearing as a function of the physical
voltage $\Phi$. The most striking influence of $\Phi$ is a  change of
the offset of the curves $D_0$ and $M_0$.
The offset is changed monotonically as a function of $\Phi$ and cannot
explain features such as dips and peaks which are found in the analytically
continued data (cf.~next section).
This is the very reason of our claim that low- to intermediate-energy scales with respect 
to $\varphi_m$ rather directly translate into low- to intermediate-energy scales with respect to $\Phi$,
although $\varphi_m$ has no direct physical meaning itself.
\begin{figure*}[htb]
\subfloat[$U=8\Gamma$, $\beta\Gamma=40$, $\mu_B B =0.02\Gamma$]
  { \includegraphics[width=0.48\linewidth]{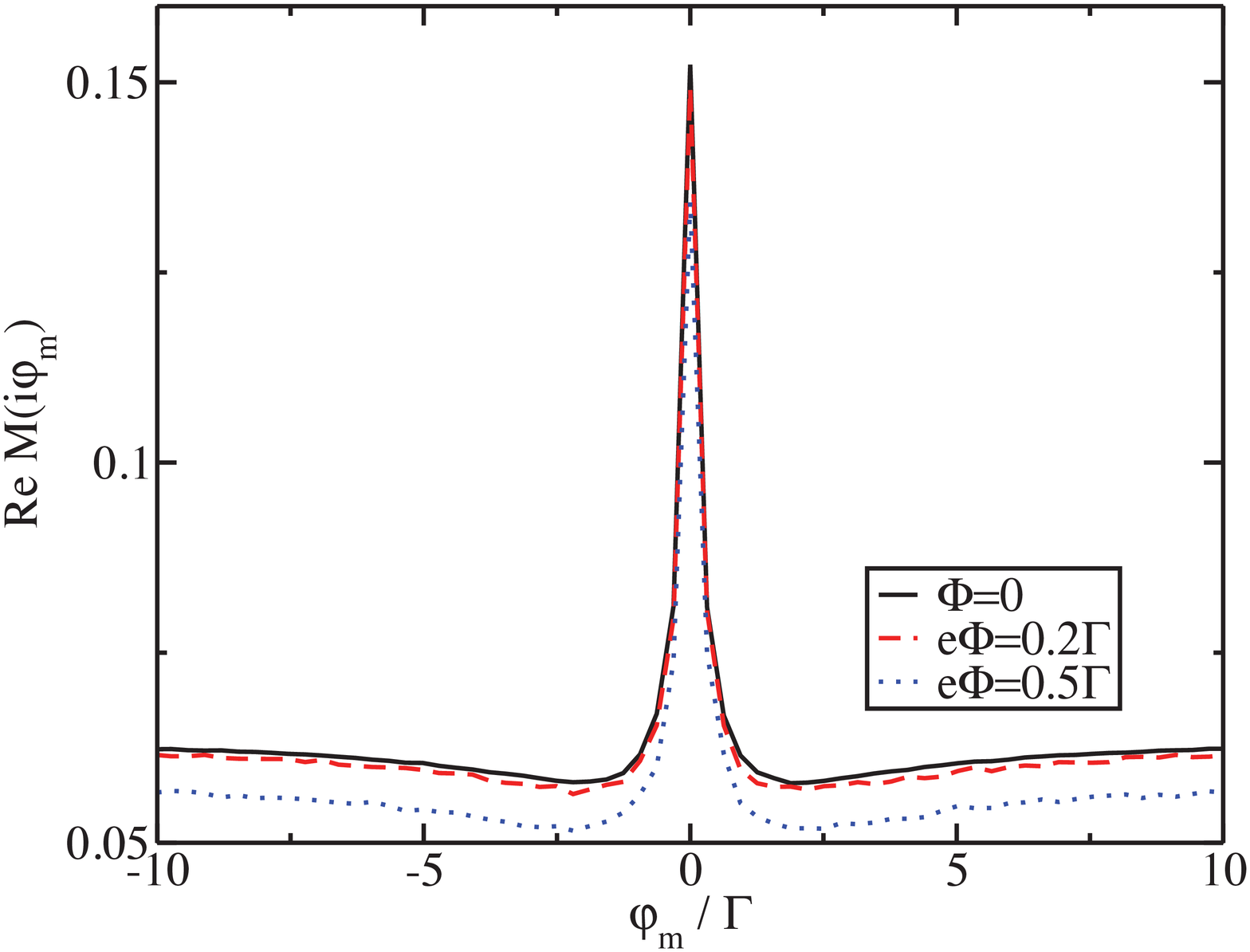}
    \label{subfig:rawdatamagnU8beta40B0.02} }
\subfloat[$\beta\Gamma=40$, $\mu_B B =0.02\Gamma$, $e\Phi=0.5\Gamma$]
  {
\includegraphics[width=0.48\linewidth]{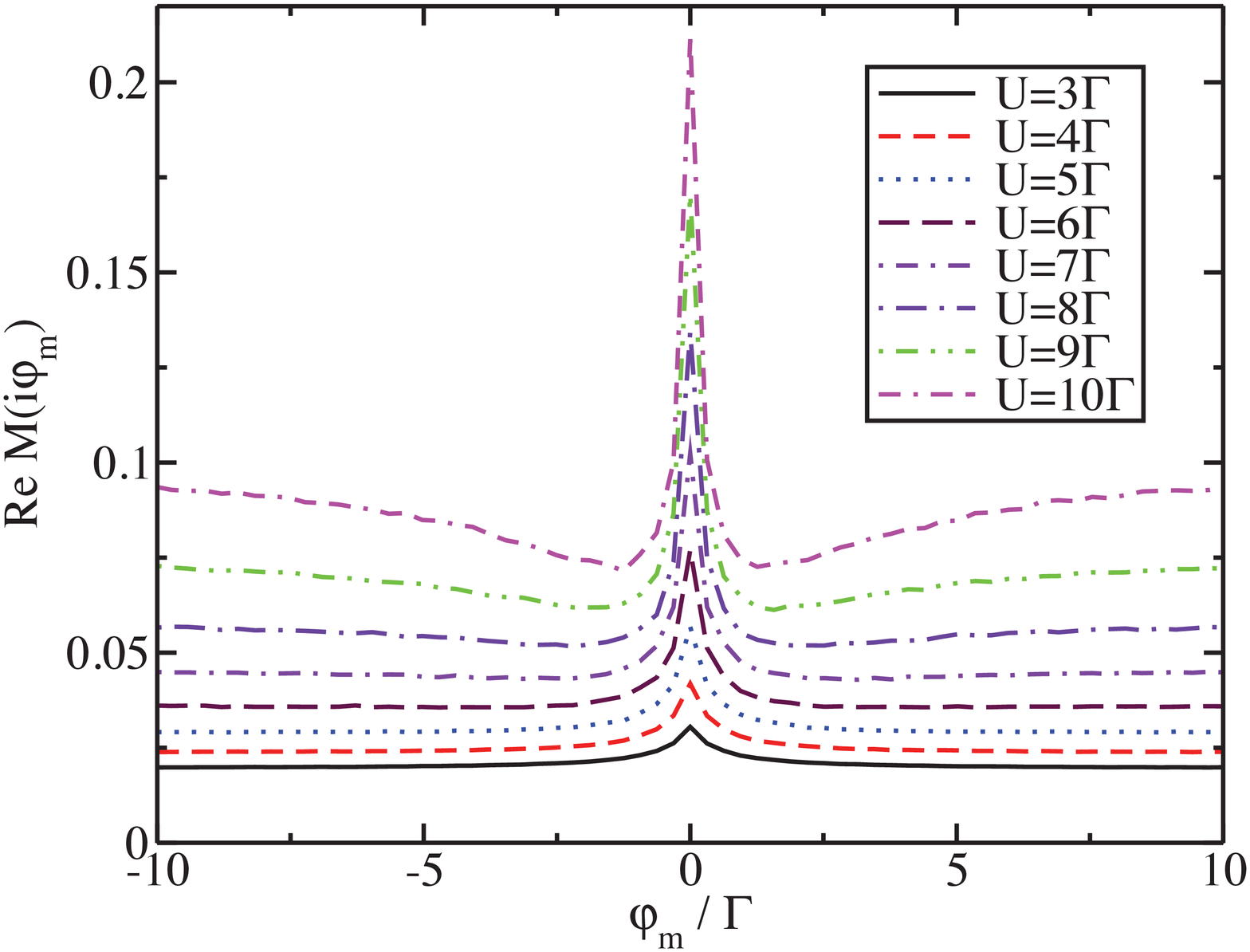}
    \label{subfig:rawdatamagndiffU}
    }
\caption{(color online) Real part of the effective-equilibrium magnetization as a function
of the Matsubara voltage.}
\label{fig:rawdatamagn}
\end{figure*}

Let us substantiate the above statement by the  data plotted in Figs.~\ref{fig:rawdatadoubleocc} and
\ref{subfig:rawdatamagnU8beta40B0.02}. 
In Fig.~\ref{fig:rawdatadoubleocc}, effective-equilibrium double occupancy curves are 
shown over a wide range of values of the physical voltage and  Coulomb interaction. 
Each curve exhibits a dip at $\varphi_m=0$. As already pointed out above, the dependence on $\Phi$ is rather
mild, except for the offset.  The same behavior is observed for the magnetization in Fig.~\ref{subfig:rawdatamagnU8beta40B0.02}, i.e.\ 
the voltage $\Phi$ merely introduces an
overall shift and a moderate smoothening of the structures.
\begin{figure*}
\subfloat[$U=5\Gamma$]{
\includegraphics[width=0.31\linewidth]{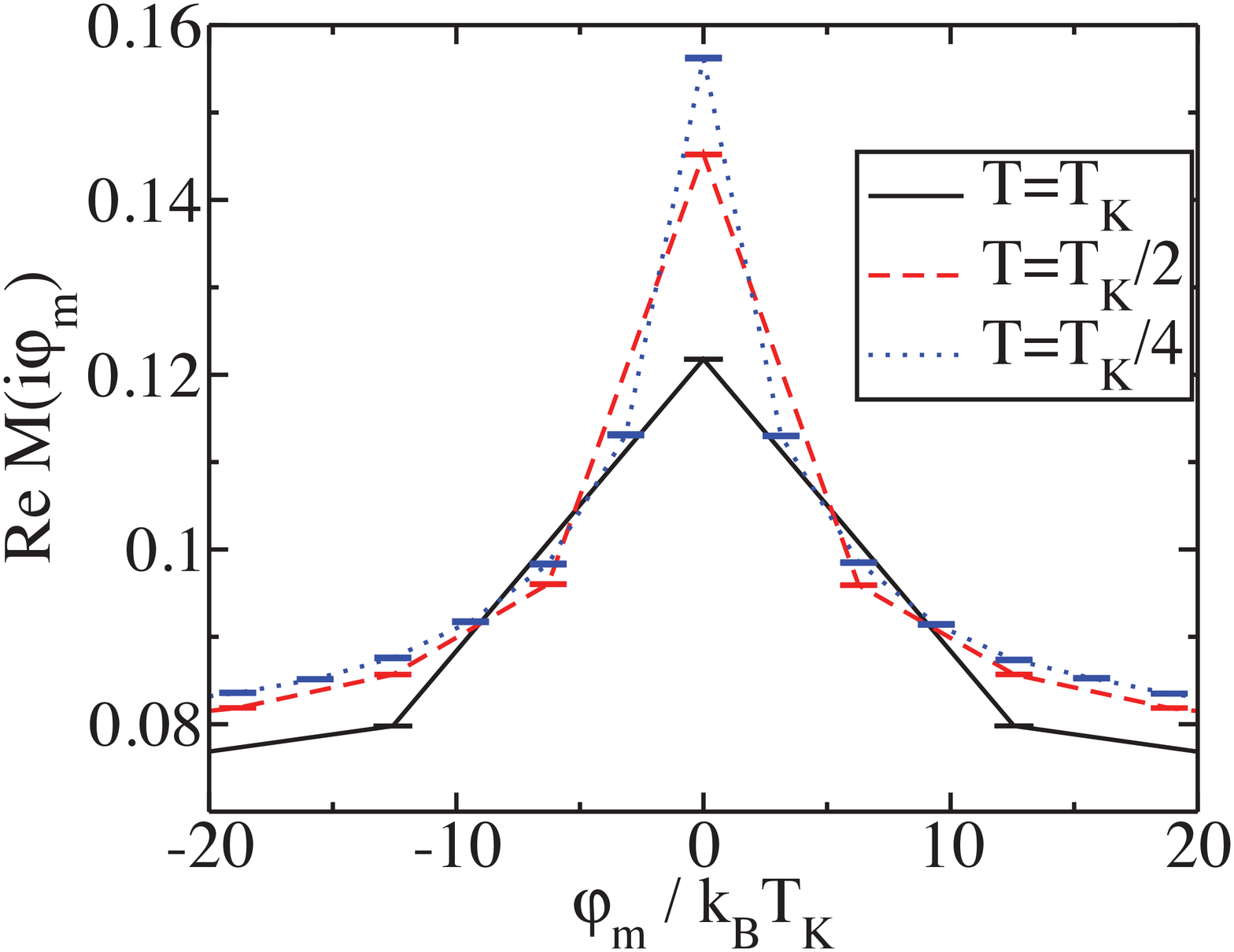}
}
\subfloat[$U=8\Gamma$]{
\includegraphics[width=0.31\linewidth]{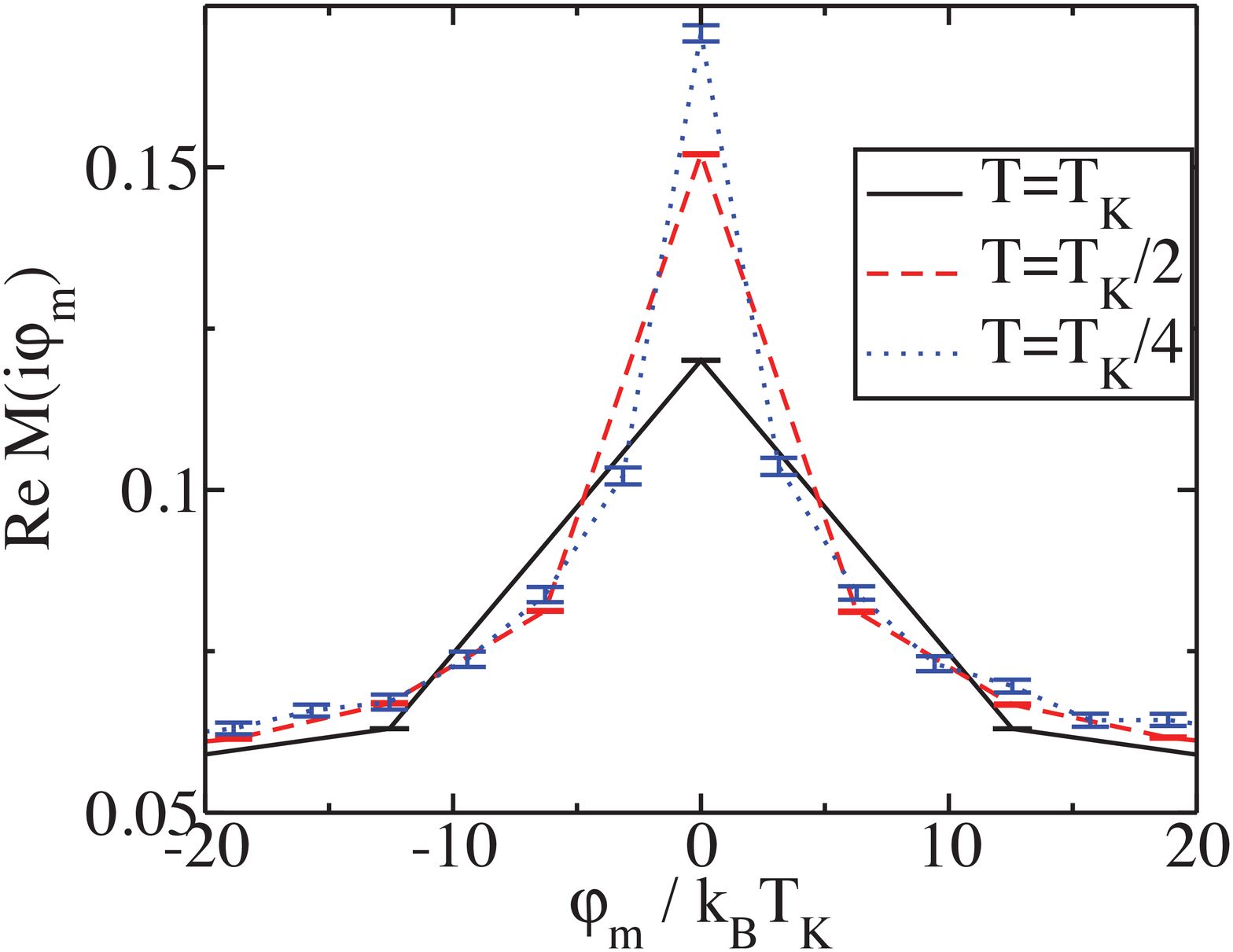}
\label{fig:KondoScalingU8}
}
\subfloat[$U=10\Gamma$]{
\includegraphics[width=0.31\linewidth]{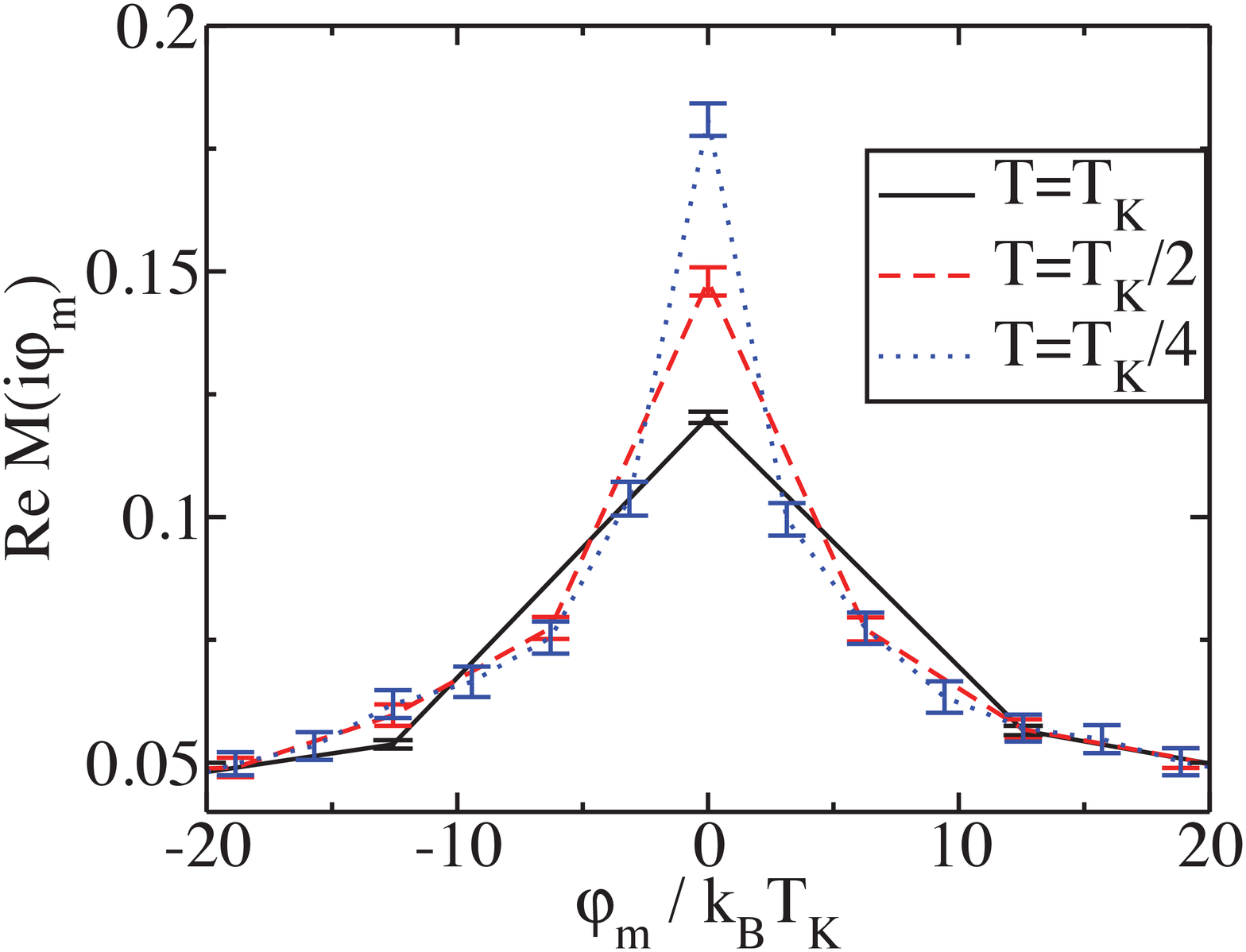}
}
\caption{(color online) Kondo scaling analysis of effective-equilibrium magnetization data at 
$\mu_B B = T_K/2$, $e\Phi =T_K/4$.
The analysis makes use of the equilibrium Kondo temperatures
$k_B T_K(U=5\Gamma)\approx \frac{1}{10}\Gamma$, 
$k_B T_K(U=8\Gamma)\approx \frac{1}{20}\Gamma$, 
$k_B T_K(U=10\Gamma)\approx \frac{1}{40}\Gamma$. 
{The latter ratios are chosen to be approximately
identical to the results of Haldane's scaling
formula.\cite{haldane}}}
\label{fig:KondoScaling}
\end{figure*}

\paragraph{Limiting behaviour $\varphi_m\to\pm\infty$.}
For each $U$ and $\Phi$ a different limit $D_0$ is obtained as
$\varphi_m\to\infty$. If the values $\beta$, $U$, $\Phi$, and in
particular $\varphi_m$ are large, the effective-equilibrium QMC simulations
start to suffer from a significant sign problem. This may
result in particularly noisy tails such as the ones for the data with
largest $\Phi$ in figure~\ref{subfig:rawdatadoubleoccU10}. In these cases, the
estimate of $D_0$ is subject to much uncertainty and limits the statistical accuracy of 
physical expectation values. 

\paragraph{Dependence on $U$.} As $U$ is increased, the depth of the dips 
in the double occupancy curves also increases. On the other hand, neither the width nor the
shape change significantly. In particular, the emergence of a Kondo scale $T_{\rm K}$
cannot be inferred from these data.  
Interestingly, for small $U$, the relative contribution of the constant term $D_0$ is large
compared to the height of the peak which emerges around $\varphi_m\approx 0$. As the
interaction increases, the central peak becomes more pronounced, and the physical
expectation value increasingly depends on the structure of the peak.

For the magnetization 
in
Fig.~\ref{subfig:rawdatamagndiffU}, a similar picture seems to emerge at first glance, namely
a strong increase of the offset $M_0$ with $U$ together with a more pronounced peak structure
at $\varphi_m=0$. The strong increase of both is readily understood as with increasing $U$
the system forms a local moment which is aligns with the external field.

\paragraph{Kondo effect.} Up to now there seems to be no evidence whatsoever for the
presence of the Kondo scale $T_{\rm K}$ in the data presented so far. On the other hand, the generation of
this many-body scale is usually considered as crucial test for any method proposed for studying
the Anderson impurity model. As already pointed out, it is quite apparent from the data in Figs.~\ref{fig:rawdatadoubleocc}
that $T_{\rm K}$ obviously does not appear to be relevant for this quantity; a fact that is already well known 
in equilibrium. There the scale $T_{\rm K}$ shows up only in a very indirect way as renormalization
of the zero temperature value respectively the scale regulating the approach to it.\footnote{F.B.~Anders, private communication.}

The situation is different for the magnetization. Here, the Kondo scale plays a crucial role 
\cite{hewson} as it determines the field-strength necessary to break up the Kondo singlet. 
Hence it must show up in the magnetization; in particular, one must actually expect a scaling
behavior with $T_{\rm K}$ for small enough fields. Let us therefore plot the magnetization as function
of Matsubara voltage in the form $M(\varphi_m/T_{\rm K})$ for values of $U$ beyond the weak-coupling
regime for fields and voltages much smaller that the corresponding equilibrium Kondo scales.
The result is shown in Fig.~ \ref{fig:KondoScaling}. Evidently, 
the width of the peak in the effective-equilibrium magnetization data is nicely scaling with the equilibrium Kondo temperature, i.e.\ 
for different values of $U$ the peak structure is essentially left invariant at fixed
values of $B$, $\Phi$, and $T$.

\subsection{Results for real voltages}
In this section we will introduce the MaxEnt procedure used to infer the
spectral functions $\varrho_{D}(\varphi)$ and $\varrho_{M}(\varphi)$ from the
effective-equilibrium QMC data. Based on this analytical continuation, we then will
discuss the physical results obtained from the auxiliary Matsubara voltage data.
\subsubsection{MaxEnt procedure}
Based on the effective-equilibrium data and the exact relation
\eqref{eq:representationStaticObservables}, it is in principle possible to
uniquely reconstruct the spectral function $\varrho_A(\varphi)$ and the
offset $\langle \hat A\rangle_\text{const}$. This is almost completely analogous to the conventional
Wick rotation. 

However, because in practice a finite set of data is considered, the
inversion of equation \eqref{eq:representationStaticObservables} is no longer
unique. On top of this, the quantum Monte-Carlo data are not exact but merely
Gaussian random variables. One may easily verify that the noise associated to the 
variables is amplified by the inversion of equation
\eqref{eq:representationStaticObservables}. As a consequence, it will always
be possible to find qualitatively very different
functions $\varrho_A(\varphi)$ which are in agreement with the QMC data. In
particular, these functions will yield physically different predictions via 
equation \eqref{eq:physExpVal}. The problem to obtain physical results from
the effective-equilibrium data is thus \emph{ill-posed}.

Since essentially the same integral equation \eqref{eq:representationStaticObservables}
also relates imaginary-time and real-time properties of conventional Green's
functions, this issue is well-known to the community.\cite{mem} Although no solution to the
problem can be provided, Bayesian inference provides a framework to
systematically incorporate a-priori information about a quantity into an
estimate. The estimate is most likely with regard to the prior information at
hand. The resulting method is called Maximum Entropy (MaxEnt).\cite{mem}

Let us consider the situation in which the offset $\langle \hat A
\rangle_\text{const}$ has
already been determined via a least-square fit. Via error propagation it has
been possible to determine the covariance matrix of the quantity
$\langle{\hat A}\rangle-\langle{\hat A}\rangle_\text{const}$, i.e. the
imaginary-voltage values of the quantity $\chi_A(z_\varphi)$ in equation
\eqref{eq:formalstrucObs}. The remaining task of the MaxEnt is to infer the spectral
function $\varrho_A(\varphi)$.
Let us furthermore assume that the data have been sufficiently transformed
with a shift function, such that the function
\begin{equation}
\varrho'_A(\varphi)=\varrho_A(\varphi)-\varrho_\text{shift}(\varphi)
\label{eq:funcshift}
\end{equation}
 is
positive (see section \ref{subsec:theobackgrnd}).

The default model for $\varrho'_A(\varphi)$ is then a positive definite
function which in principle should contain features which determine in 
particular the high-energy behaviour, if known.\cite{mem} In the case of
Green's functions, perturbation theory or higher-temperature solutions often
give good default models.\cite{mem}
In our case, apart from that we used a shift function to construct the
positive spectrum, nothing is known about the function, so a flat default
model is preferable. As consequence, we use the shift function itself as the
default model in the actual computation. For simplicity, let us call the
to-be-inferred spectrum $\varrho(\varphi)$ and the default model
$\varrho_\text{def}(\varphi)$.

On the one hand, the default model gives rise to a relative entropy
\cite{mem}
\begin{equation*}
S = \int \Dfrtl\varphi\left[\varrho(\varphi) - \varrho_\text{def}(\varphi) 
- \varrho(\varphi) \log \frac{\varrho(\varphi)}{\varrho_\text{def}(\varphi)}\right]
\end{equation*} 
of the spectral function. 
On the other hand, the (transformed) effective-equilibrium simulation data 
with mean values $\bar a_i$ and covariance $C_{ij}$ yield the measure
\begin{equation}
\chi^2 = \frac{1}{2} \sum_{i,j}^{N_{QMC}} 
(\bar a_i - y_i) C^{-1}_{ij}(\bar a_j - y_j).
\end{equation}
for the quality of the fit.
Here $y_i$ are the fit values which result from transforming the considered $\varrho(\varphi)$
to the data space, and $N_{QMC}$ is the number of QMC data points $\bar a_i$. Within the MaxEnt 
it follows that a functional $Q=\chi^2-\alpha S$ must be minimized, where $\alpha >0$ is some
hyperparameter.\cite{mem}
 
In order to determine $\alpha$, there are several methods, for example the ``historic'' and 
the ``classic'' MaxEnt.\cite{mem} The former extracts information from the
Monte-Carlo data up to the point at which the $\chi^2 = N_{QMC}$,
i.e.~the MaxEnt regularization parameter is fixed to the value at which
$\chi^2 = N_{QMC}$. 
The latter (``classic'' MaxEnt) extracts information from QMC data to a larger extent. Based on
the probability distribution implied by the default model and
maximum-likelihood functionals, a posterior probability of the MaxEnt
regularization parameter $\alpha$ is maximized. Because information from the
default model is again incorporated rather explicitly, this strategy is
particularly good for default models which are close to the actual solution.
A rather general feature of ``classic'' MaxEnt appears to be that the
$\chi^2$ value of the inferred estimate is generally much smaller than the
``historic'' value of $N_{QMC}$. Our feeling is that this aspect makes the ``classic'' estimate
more sensitive to statistical fluctuations and vulnerable for over-fitting, but on the
same side, the estimate is less biased. A similar increase in fluctuations
was pointed out in a recent study.\cite{gunnarsson}
At least if Bayesian evidence coming from the data is weak, the ``historic'' 
MaxEnt, on the other hand is more biased towards the default
model value, since its estimate is more conservative with regard to the
$\chi^2$.
In our case, the default-model estimate is given by the constant
offset $D_0$, because our default models are chosen to be even functions with
respect to $\varphi$.

\begin{figure*}[htb]
\subfloat[inferred MaxEnt spectra $\varrho_D'(\varphi)$]
{
\includegraphics[width=0.48\linewidth]{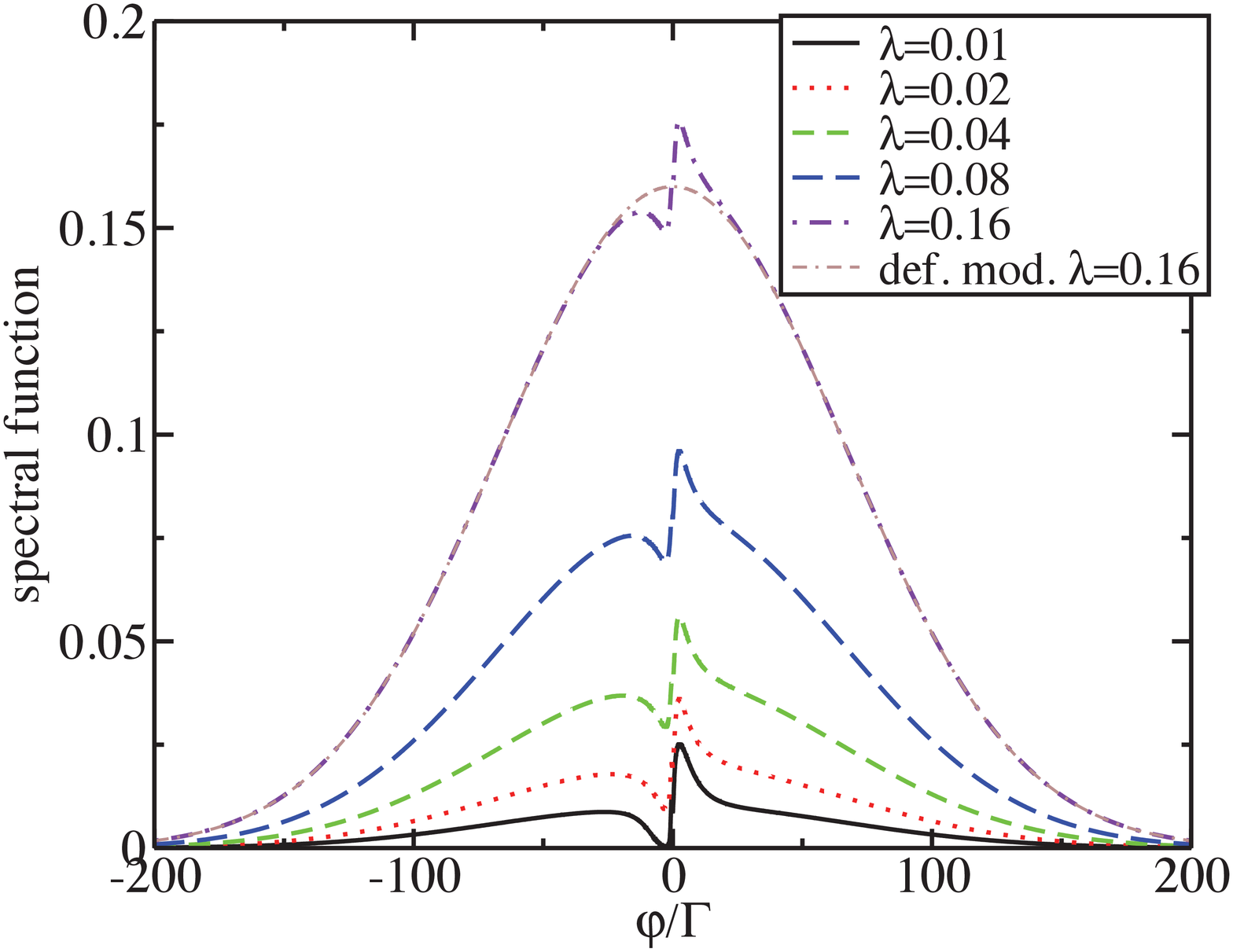}
}
\subfloat[resulting spectral functions $\varrho_D(\varphi)$]
{
\includegraphics[width=0.48\linewidth]{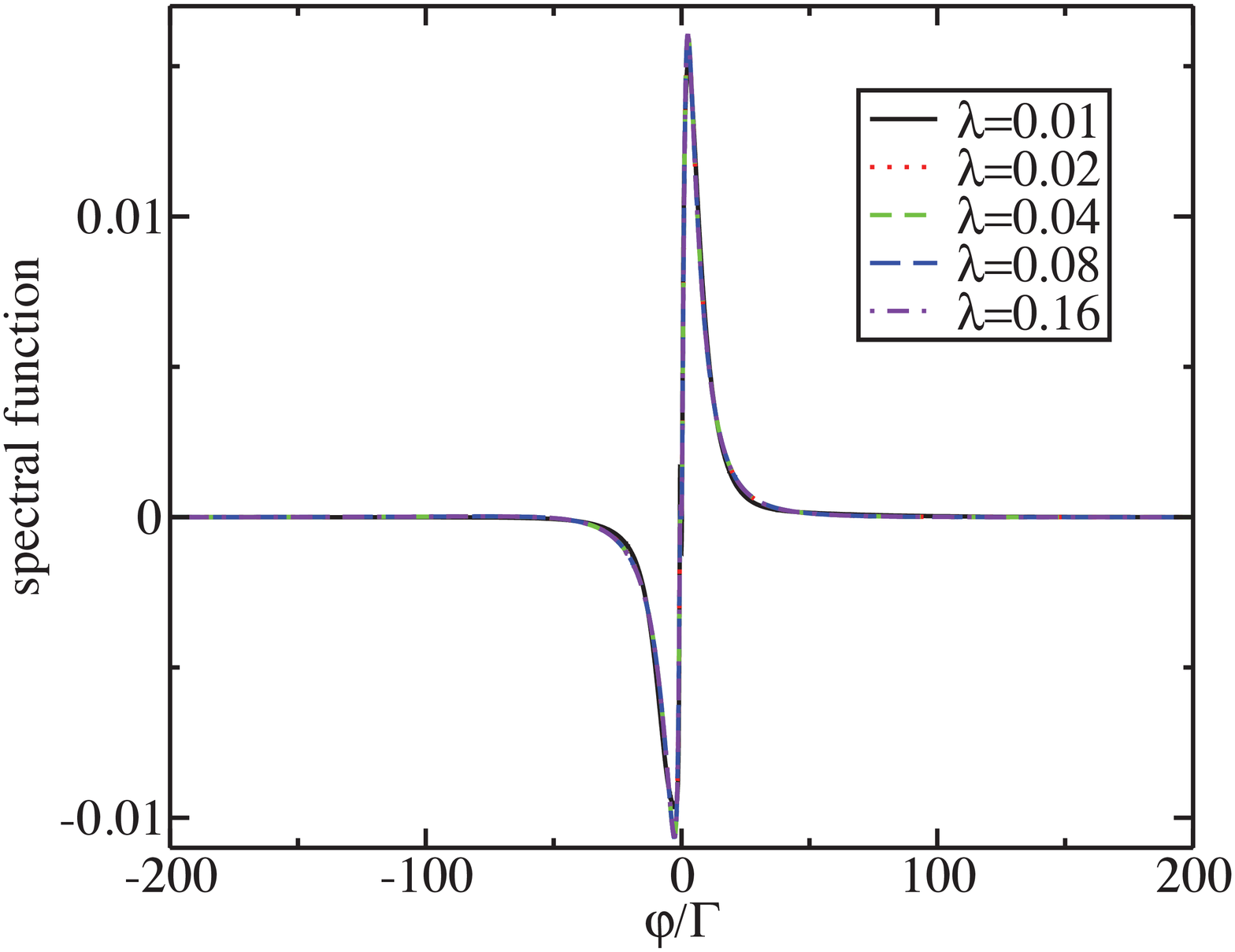}
}
\caption{(color online) MaxEnt inference process for the double occupancy. Parameters are
$U=5$, $e\Phi=0.25\Gamma$, $\beta=20\Gamma^{-1}$. 
{
Due to lack of prior
knowledge, we use a flat default model, i.e.~the shift function
$\varrho_\text{shift}(\varphi)$, see Eq.~\eqref{eq:shift_function}.
Remember that the actual spectral function $\varrho_D(\varphi)$ was shifted
to a positive one, $\varrho_D'(\varphi)$, via
equation \eqref{eq:funcshift}.
One finds that the different equivalent ways of imposing a flat default model
for $\varrho_D(\varphi)$ yield practically the same spectral function.
Nevertheless, computing the physical value \eqref{eq:physExpVal} yields
values which are distributed over a certain range. This range is displayed as
error bars in the results plots Figs.~\ref{fig:diffUdoccvsPT} and
\ref{fig:magnsusceptibility}.
}
}
\label{fig:maxent}
\end{figure*}

As shift functions, wide Gaussians with width $\sigma=\frac{200}{3}\Gamma$ were used,
i.e.
\begin{equation}\label{eq:shift_function}
\varrho_\text{shift}(\varphi) = \lambda \cdot
\euler{-\varphi^2 /2\sigma^2}.
\end{equation} 
The amplitude of the functions was varied in such a way that positive functions
could be inferred. The different values for differently scaled functions give
rise to a certain interval of expectation values, which will be plotted as a
result, in the following.
An example for the set of inferred functions obtained for a single
non-equilibrium system is shown in figure \ref{fig:maxent}.
The left panel shows the actually performed MaxEnt for the shifted spectral
functions, using ``historic'' MaxEnt. Resulting from a flat default model for the 
function $\varrho_D(\varphi)$,
the shift function acts as default model here. 
In this case, choosing a parameter $\lambda < 0.01$ yields
artifacts in the physical solutions, because the negative regions of
$\varrho(\varphi)$ cannot be represented any more.
The corresponding actual
spectral functions $\varrho(\varphi)$, obtained by subtracting the shift function
(\ref{eq:shift_function}) from the data in the left panel, are shown in the right panel of
Fig.~\ref{fig:maxent}. 
{
The flat default model represents our lack of prior information
about the solution and the preference of a smooth solution in case of
uncertainty. In general, the different realizations of a flat default model
with the shift functions yields almost but not exactly the same spectral
functions.
In case of limited QMC data quality, it is well known\cite{mem} that the usage of a flat default model
yields less accurate spectra than an appropriately constructed more informative default model.
For example, in case of conventional equilibrium spectral functions of Fermi or Bose systems, a
default model should preferably obtain the correct low-order moments, which can often be
computed exactly.
It can thus be expected that quantities that are calculated from the spectra
inferred using the flat default model are biased towards a certain value.
Nevertheless, an increase in data quality will eventually reduce the bias of
the estimated quantity.
We also expect that the precision of our method can be increased by
the development of default models which contain additional information like moments. However,
at present such type of information is not yet available.
}

In order to obtain a rough estimate on the error of a physical estimate, 
we will plot the intervals which are generated by computing the estimates for
different values of $\lambda$. Typically, a range from $\lambda=0.01$ to
$\lambda=0.16$ is imposed, unless the negative regions of
$\varrho(\varphi)$ cannot be represented. For the magnetic susceptibility,
the same strategy is used.

\subsubsection{Double occupancy}
\begin{figure}
\includegraphics[width=\linewidth]{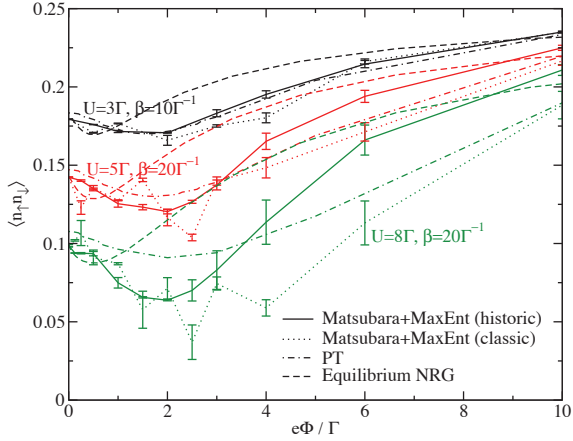}
\caption{(color online) Double occupancy as a function of the bias voltage at different
values of $U$, as compared to second-order perturbation theory. In addition, the dashed lines
show the temperature dependence of $\langle n_\uparrow n_\downarrow\rangle$ in equlibrium 
as obtained by NRG, 
{assuming $e\Phi=k_{\rm B}T$ (see text).}}
\label{fig:diffUdoccvsPT}
\end{figure}

We will now discuss the analytically continued data of the double occupancy
and compare it with respect to zero-temperature second-order perturbation
theory.\cite{muehlbacher}
In figure \ref{fig:diffUdoccvsPT} we show double occupancy data for different values of 
the Coulomb interaction computed with the 
two different MaxEnt estimators. 

The complementary behaviour of the two estimators may be well observed in
Fig.~\ref{fig:diffUdoccvsPT}. In the large-bias limit, in which the
perturbation theory  may
be expected to be correct, the classic estimator is closer, and the historic
estimate is systematically too high. This is in agreement with our
expectation that the historic estimate will be biased from above in case of
rather weak Bayesian evidence from QMC data, because the ill-posed continuation problem
is particularly severe at high energies.\cite{mem}
Apart from some fluctuations in the
``classic'' estimator, the same curves are predicted for small voltages.
It is important to note that error bars in the figures do not denote statistical errors (which
cannot be estimated), but the range of values which a given set of 
symmetric default models generates. 

As compared to the second-order perturbation theory, we find that both
methods agree perfectly for interaction strength $U=3\Gamma$. Also both methods
predict a minimum in the double occupancy at voltage $e\Phi\approx2\Gamma$ which slowly shifts
to larger values of $\Phi$ and becomes
increasingly distinguished as the interaction is increased. There is, however, a clear
difference concerning the magnitude of this minimum, which appears much more pronounced
in the QMC data as in the perturbation theory. Note that this seems to be the case for both MaxEnt estimators. At present the origin of the deviation is not clear.

One of the issues related to the $\Phi$ dependence of stationary non-equilibrium quantities is
to what extent they can be mapped onto an effective equilibrium temperature dependence. To have an idea
whether this mapping works, we included in Fig.~\ref{fig:diffUdoccvsPT} also the
corresponding curves for  $\langle n_\uparrow n_\downarrow\rangle(T)$ as obtained from
an NRG equilibrium calculation, assuming $e\Phi=k_{\rm B}T$. Quite apparently, the values at $\Phi\to0$ nicely coincide, which
also tells us that the Matsubara voltage QMC reproduces the proper low bias results even for
strong coupling. Note that perturbation theory here deviates systematically with increasing $U$.
However, the dependence of $\langle n_\uparrow n_\downarrow\rangle(\Phi)$ cannot be 
mapped even qualitatively onto $\langle n_\uparrow n_\downarrow\rangle(T)$ by a simple ansatz
$\Phi\hat{=}\alpha\cdot T$ with some value $\alpha$ for any of the $U$ values considered here.
From this observation we would thus conclude that such a mapping is -- at least for the simplest possible quantity -- not appropriate.

\subsubsection{Magnetic Susceptibility}
Similarly, the magnetic susceptibility may be computed as a function of the
bias voltage by analytical continuation of the QMC data. As an example, we show the
result for  $U=8\Gamma$ at the temperature $T=T_K/2$ 
and magnetic field $\mu_BB=k_BT_K/2$
\begin{figure}
\includegraphics[width=\linewidth]{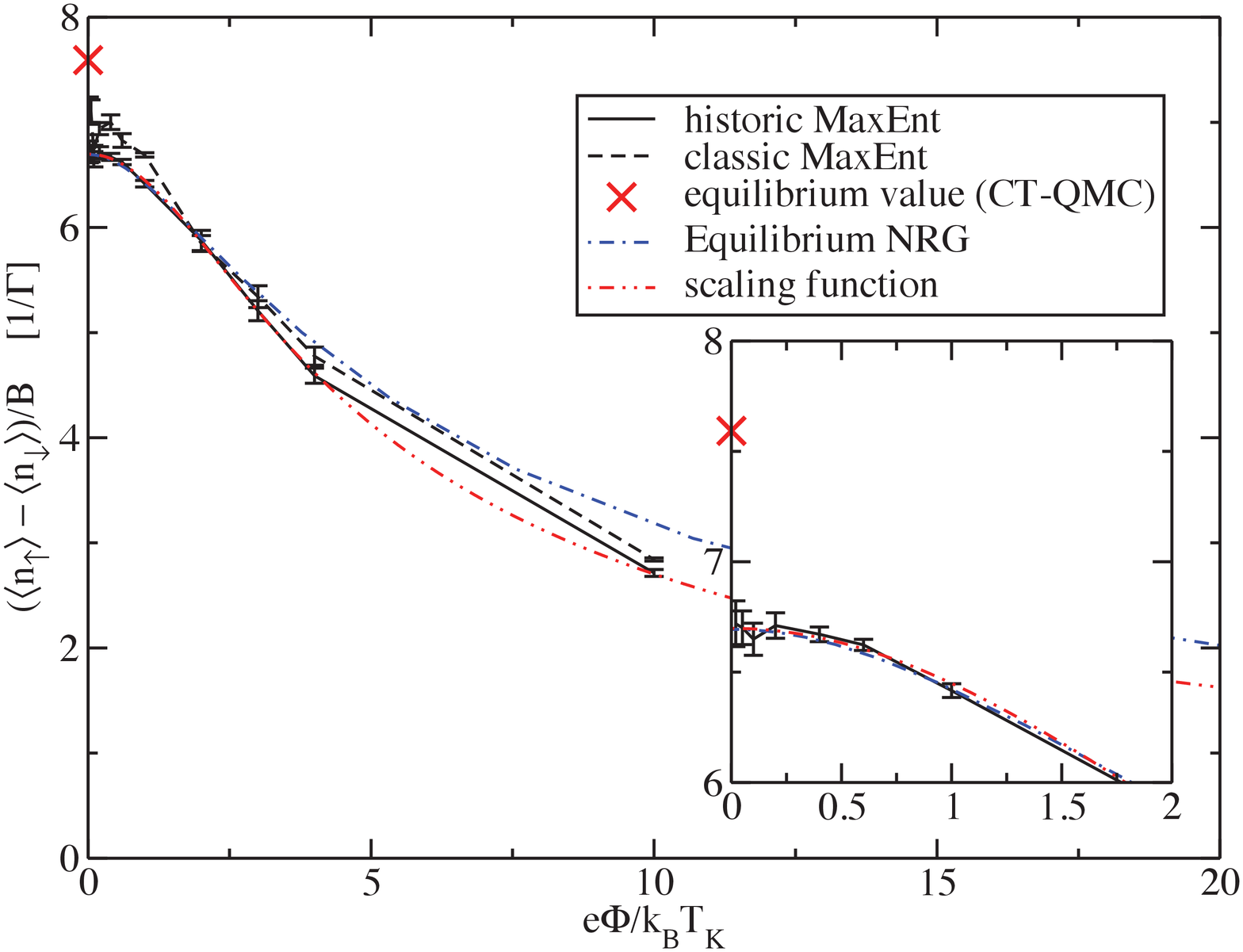}
\caption{(color online) Magnetic susceptibility as a function of  bias voltage in the
Kondo regime  $U=8\Gamma$ at $\mu_B B =k_B T_K/2$, $T=T_K/2$. The dot-dashed line
represents an equilibrium NRG calculation for $T\ge T_{\rm K}/2$, rescaled in both magnitude 
and temperature to match the low-bias behavior 
{of historic
MaxEnt} (see inset). The
double-dot-dash curve 
finally is a
fit 
{of historic MaxEnt} to some scaling function (see text).}
\label{fig:magnsusceptibility}
\end{figure}
in Fig.~\ref{fig:magnsusceptibility}. 
When we compare our continuation results 
at $\Phi\to0$
{
to the exact low-bias limit (i.e.~the equilibrium value, displayed as a cross
in Fig.~\ref{fig:magnsusceptibility}), the historic MaxEnt is again
more strongly biased than the classic MaxEnt, i.e.~the deviation from the
equilibrium value is stronger. With insufficient QMC information, the outcome is more biased
towards the flat default model and from Eq.~(\ref{eq:physExpVal}) the
integral vanishes in such limit. The constant offset $M_0$ lies below the actual
physical limit, and therefore, as QMC quality improves, our estimate approaches the
correct limit from below.
}
  Again, the classic MaxEnt is subject to stronger fluctuations. 

In physical terms, the decay in magnetic susceptibility is because of the
destruction of the Kondo effect due to the decoherence introduced by the
bias voltage. This is in principle similar to the equilibrium behaviour found as a function of
temperature.\cite{hewson} The scale on which the decay of the magnetization
takes place appears to be already visible within the imaginary-voltage data
shown in Fig.~\ref{fig:KondoScalingU8}. Apparently, this is due to the
rather weak voltage-dependence of imaginary-voltage data (cf.~figure 
\ref{subfig:rawdatamagnU8beta40B0.02}). Voltages above $10k_BT_K$ were not
accessible to the MaxEnt, due to a strong sign problem occurring for the QMC
simulations of the effective-equilibrium systems associated to the high-$\varphi_m$ 
tails.

We again may compare the voltage dependence of the stationary non-equilibrium magnetization to the temperature dependence in equilibrium. Since we here are at a finite temperature $T=T_{\rm K}/2$, 
hence the magnetization is smaller than the value at $T=0$, the natural
thing to look at is the curve $M(T)\cdot [M(T_{\rm K}/2)/M(0)]$ and
rescale temperature with an appropriate factor. The result is shown as
dot-dashed line 
in Fig.~\ref{fig:magnsusceptibility}. Although one
can reach a reasonable match for low voltages, a significant deviation occurs already at 
moderate bias. Thus there does not seem to exist a simple mapping $\Phi\to T$ which will bring the
curves to overlap, .i.e.\ it again seems doubtful that one can describe the effect of finite bias voltage
by an effective temperature scale, at least beyond small bias voltages of the order of the Kondo scale.

On the other hand, a rather good account for all data can be achieved by the very simple ansatz
\begin{eqnarray*}
\frac{m(\Phi)}{B}&\approx&\frac{a}{B}
\frac{1}{\frac{\tilde{\Phi}^2}{\sqrt{b^2
+\tilde{\Phi}^{2}}}+c}
\end{eqnarray*}
where $\tilde{\Phi}:=\Phi/(2T_{\rm K})$. The result of this fit with
$a=0.52$, $b\approx2$ and $c\approx3$
 is shown as the double-dot-dash curve in Fig.~\ref{fig:magnsusceptibility}. Note that this formula gives the right behavior in the two limits
$\Phi\to0$, viz $M/B\propto1-c\tilde{\Phi}^2$ with some numerical constant $c$, and $\Phi\to\infty$, viz $M/B\propto 1/\Phi$. From scaling analysis\cite{rosch} one would expect that, in particular for
large bias, additional logarithmic corrections appear. Due to the limited data space available we are
of course not able to resolve those; furthermore, it is not clear if these logarithmic corrections will
actually be visible in the intermediate coupling regime studied here, due to residual charge 
fluctuations. We therefore view the above formula as a reasonable
description in the regime of bias, temperature and field of the order of the Kondo temperature for
the intermediate coupling regime of the SIAM. 

\section{Summary}\label{sec:V}
The present paper presents a detailed study on how the imaginary-voltage
formalism proposed in Ref.\ \onlinecite{prl07} relates to Keldysh theory. Using series resummations, we are able to show up to all orders 
that static expectation values of observables, which satisfy certain symmetry relations with respect to the Keldysh contour,
map exactly onto the corresponding expressions in Keldysh perturbation theory.
In particular, it was pointed out that in order to obtain a physical
expectation value, the limiting process $\imag\varphi_m\to\Phi$ has to be
taken as principal-value. This prescription ensures, that one generates the principal-value integrals which
emerge in the proper real-time theory. For dynamical correlation functions, this was
shown explicitly up to fourth order of perturbation theory.

As one important novel result of the present paper we were able to 
{provide}
an exact spectral representation for static
expectation values similar to a Lehmann representation.
Based on the representation, using unbiased numerical data from continuous-time 
quantum Monte-Carlo simulations, we found that the evaluation of the 
limiting procedure as principal-value expression does indeed give real numbers as physical expectation values. Consequently, the
theory is found to be fully consistent in this respect beyond the perturbation arguments given. 
The double occupancy as function of bias voltage computed this way shows features similar to straight-forward second-order perturbation
theory, but we find them to be more pronounced. For the magnetic
susceptibility we were able to give numerical estimates on the destruction of
the Kondo effect. A comparison to equilibrium NRG shows that the dependence on bias 
voltage for both, the double occupancy and the magnetic susceptibility, cannot be explained by a simple
effective-temperature interpretation.

\section{Acknowledgments}
We acknowledge valuable discussions with M.~Jarrell, J.~Freericks, F.B.~Anders, S.~Schmitt,
K.~Sch\"onhammer, A.~Schiller. AD 
acknowlegdes financial support by the DAAD through the PPP exchange program, 
and JH acknowledges the National Science Foundation with the
Grant number DMR-0907150, TP the German Science Foundation through SFB 602.
AD and TP would also like to acknowledge
computer support by the HLRN, the GWDG and the GOEGRID initiative of the University
of G\"ottingen.
Parts of the implementation are based on the ALPS 1.3 library \cite{alps}.

\appendix

\section{Cancellation of overlapping $\delta$-functions in
Eq.~(\ref{eq:double})}
\label{app:double}
With a set of $\{\psi^\dagger_{\alpha_i
k_i\sigma_i},\psi_{\alpha_i k_i\sigma_i};i=1,\cdots,6\}$ appearing for
the matrix elements in Eq.~(\ref{eq:double}), we categorize the thermal
factor $e^{\beta\Phi Y_{0\{n,m,l\}}}$ as follows. (i) If $Y_{0n}=Y_{0m}=Y_{0l}$,
Eq.~(\ref{eq:double}) vanishes. (ii) If only one of $Y_{0n},Y_{0m},Y_{0l}$ is
different from others, $(Y_{0n},Y_{0m},Y_{0l})\in\{(Y_0,Y_0,Y_0+1),
(Y_0+1,Y_0,Y_0),(Y_0,Y_0+1,Y_0),(Y_0,Y_0,Y_0+2),(Y_0+2,Y_0,Y_0),(Y_0,Y_0+2,Y_0)\}$
for some reference value $Y_0$. If we take the case of
$(Y_{0n},Y_{0m},Y_{0l})=(Y_0,Y_0,Y_0+1)$, the terms contributing for the
matrix elements $V_{nm}$, $V_{ml}$ and $A_{ln}$ are from
$\psi^\dagger_{\tilde\alpha_1 \tilde{k}_1}\psi^\dagger_{\tilde\alpha_2
\tilde{k}_2}\psi_{\tilde\alpha_3
\tilde{k}_3}\psi_{\tilde\alpha_4 \tilde{k}_4}$, $\psi^\dagger_{R
k_1}\psi_{L k_2}\psi^\dagger_{\tilde\alpha_5
\tilde{k}_5}\psi_{\tilde\alpha_6 \tilde{k}_6}$, and $\psi^\dagger_{L k_2}\psi_{R
k_1}\psi^\dagger_{\tilde\alpha_7
\tilde{k}_7}\psi_{\tilde\alpha_8 \tilde{k}_8}$, respectively, where
$(\tilde{k}_1,\cdots,\tilde{k}_8)$ is a some permutation of
$(k_3,k_3,k_4,k_4,\cdots,k_6,k_6)$. The reservoir indices should be
chosen such that $\tilde\alpha_5=\tilde\alpha_6$ and
$\tilde\alpha_7=\tilde\alpha_8$, and
$(\tilde\alpha_1,\tilde\alpha_2,\tilde\alpha_3,\tilde\alpha_4)$ should
satisfy $Y_{0n}=Y_{0m}$. The $\tilde\alpha_i$ indices are summed over
for $L/R$. Then the term in Eq.~(\ref{eq:double}) becomes proportional
to
$$
(t_Lt_R)^2(t_L^2+t_R^2)^4\prod_{i=1,6}|g(k_i)|^2e^{\beta\Phi
Y_0}(1-2+e^{\beta\Phi}).
$$
For other combinations of $(Y_{0n},Y_{0m},Y_{0l})=(Y_0,Y_0+1,Y_0),
(Y_0+1,Y_0,Y_0)$ the thermal factor becomes $(1-2e^{\beta\Phi}+1)$ and
$(e^{\beta\Phi}-2+1)$, respectively, and all three contributions sum up to zero.
With the case of $(Y_0,Y_0,Y_0+2)$, the contribution becomes
$(t_Lt_R)^4(t_L^2+t_R^2)^2\prod_{i=1,6}|g(k_i)|^2e^{\beta\Phi
Y_0}(1-2+e^{2\beta\Phi})$. The other terms have factors of
$(1-2e^{2\beta\Phi}+1),(e^{2\beta\Phi}-2+1)$, and these sum up to zero
again.

(iii) When all of $Y_{0n},Y_{0m},Y_{0l}$ are different,
$(Y_{0n},Y_{0m},Y_{0l})$ is a permutation of $(Y_0,Y_0+1,Y_0+2)$. Since
$\hat{V},\hat{A}$ are at most two-particle operators the difference of
$Y$-values between states cannot be greater than two. If
$(Y_{0n},Y_{0m},Y_{0l})=(Y_0,Y_0+1,Y_0+2)$, the factor in
Eq.~(\ref{eq:double}) becomes proportional to
$$
(t_Lt_R)^4(t_L^2+t_R^2)^2\prod_{i=1,6}|g(k_i)|^2e^{\beta\Phi
Y_0}(1-2e^{\beta\Phi}+e^{2\beta\Phi}).
$$
Permuting $(Y_0,Y_0+1,Y_0+2)$ the sum of the thermal factors can be
easily shown to be zero.

\begin{widetext}

\section{Fourth order expansion of electron self-energy}
\label{app:fourth}

\begin{figure}
\rotatebox{0}{\resizebox{4.2in}{!}{\includegraphics{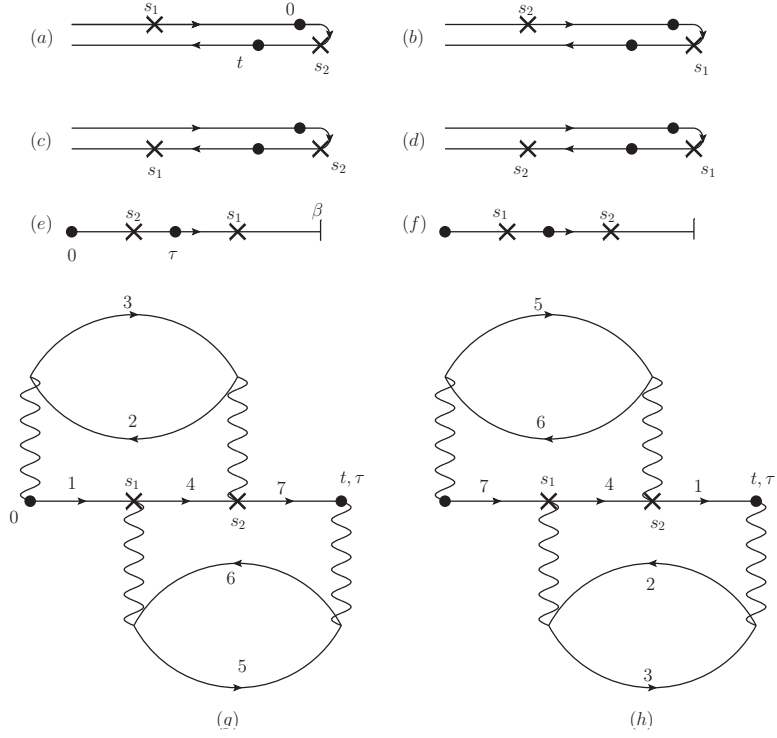}}}
\caption{\small (a-d) Real-time Keldysh contour for self-energy
$\Sigma_{(4)}^>(t,0)$ in the fourth order perturbation when one intermediate
time is in the finite interval $[0,t]$ and the other time in along the
contour stretching to $-\infty$. The Wick's contraction
is taken as shown in (g-h). The dummy label of (g) is used for
time-orderings (a,c,e) and (h) used for (b,d,f). 
The cross represents the intermediate times
$s_1$ and $s_2$ for interaction, in addition to the
creation/annihilation points $0$ and
$t$.
}
\label{fig:fourth}
\end{figure}

We investigate the energy-pole structure in the real-time perturbation
expansion to verify that the $\delta$-function residue disappears and
the energy denominators can be interpreted as principal-valued. In the
following we consider the perturbation expansion for the
self-energy in the fourth order of Coulomb parameter $U$,
$\Sigma_{(4)}^>(t,0)$ according to
the time-orderings along the Keldysh contour, FIG.~\ref{fig:fourth}(a-d).
Different types of time-orderings will be considered shortly.
These time-orderings have one of the intermediate time (marked as cross)
within a finite time-interval fixed by time at $0$ and $t$. Given a
time-ordering, a particular Wick's contraction should be chosen.
The chosen Wick's contraction is according to the diagrams in (g-h) which
correspond to the most non-trivial vertex correction.

We can evaluate each contribution as follows.
\begin{eqnarray}
S_a&=&
f_1 f_2\bar{f}_3\bar{f}_4\bar{f}_5 f_6 \bar{f}_7
\int_{-\infty}^0 ds_1\int_0^t ds_2 
e^{-i(\epsilon_1
-\epsilon_4-\epsilon_5+\epsilon_6-i\eta)s_1
-i(-\epsilon_2+\epsilon_3+\epsilon_4-\epsilon_7)s_2
-i(\epsilon_5-\epsilon_6+\epsilon_7)t}
\label{eq:sa1}\\
S_b&=&
\bar{f}_1 f_2\bar{f}_3f_4 f_5 \bar{f}_6 \bar{f}_7
\int_{-\infty}^0 ds_2\int_0^t ds_1 
e^{-i(\epsilon_2
-\epsilon_3-\epsilon_4+\epsilon_7)s_1
-i(-\epsilon_1+\epsilon_4+\epsilon_5-\epsilon_6-i\eta)s_2
-i(\epsilon_1-\epsilon_2+\epsilon_3)t}\\
S_c&=&
\bar{f}_1 f_2\bar{f}_3f_4f_5 \bar{f}_6 \bar{f}_7
\int_t^{-\infty} ds_1\int_0^t ds_2 
e^{-i(\epsilon_1
-\epsilon_4-\epsilon_5+\epsilon_6-i\eta)s_1
-i(-\epsilon_2+\epsilon_3+\epsilon_4-\epsilon_7)s_2
-i(\epsilon_5-\epsilon_6+\epsilon_7)t}
\label{eq:sc}\\
S_d&=&
f_1 f_2\bar{f}_3\bar{f}_4 \bar{f}_5 f_6 \bar{f}_7
\int_t^{-\infty} ds_2\int_0^t ds_1 
e^{-i(\epsilon_2
-\epsilon_3-\epsilon_4+\epsilon_7)s_1
-i(-\epsilon_1+\epsilon_4+\epsilon_5-\epsilon_6-i\eta)s_2
-i(\epsilon_1-\epsilon_2+\epsilon_3)t}.
\end{eqnarray}
In these shorthand notation (as discussed in the main text), 
we omitted the expression $U^4[\prod_i\int
d\epsilon_i |g_d(\epsilon_i)|^2]$ which is common to all $S_i$ terms.
$f_i=[1+e^{\beta(\epsilon_i-\alpha_i\Phi/2)}]^{-1}$ and
$\bar{f}_i=1-f_i$. After some algebra, we get
\begin{equation}
S_a+S_d
=-\frac{2f_1 f_2\bar{f}_3\bar{f}_4\bar{f}_5 f_6 \bar{f}_7}{
(-\epsilon_2+\epsilon_3+\epsilon_4-\epsilon_7)(\epsilon_1-\epsilon_4-\epsilon_5+\epsilon_6)}
[e^{-i(-\epsilon_2+\epsilon_3+\epsilon_4+\epsilon_5-\epsilon_6)t}
-e^{-i(\epsilon_5-\epsilon_6+\epsilon_7)t}].
\end{equation}
The exponential terms cancel each other at the energy poles and
$(\epsilon_2-\epsilon_3-\epsilon_4+\epsilon_7)^{-1}$ and
$(\epsilon_1-\epsilon_4-\epsilon_5+\epsilon_6)^{-1}$ give well-defined
principal-valued integral. This is a typical behavior since an integral
within a finite interval $(0,t)$ does not need the convergence factor $e^{\eta
t}$ and, accordingly, principal-valued integral is enough. The same can
be said for the combination $S_b+S_c$.

Now, we take the imaginary-time contours in FIG.~\ref{fig:fourth}(e-f).
After straightforward calculations, we have
($\tilde{\epsilon}_i=\epsilon_i-\alpha_i \epsilon_\varphi/2$)
\begin{eqnarray}
S_e &=&
f_1 f_2\bar{f}_3\bar{f}_4\bar{f}_5 f_6 \bar{f}_7
\frac{e^{-(-\tilde\epsilon_2
+\tilde\epsilon_3+\tilde\epsilon_4+\tilde\epsilon_5-\tilde\epsilon_6)\tau}
-e^{-(\tilde\epsilon_5
-\tilde\epsilon_6+\tilde\epsilon_7)\tau}}{
(\tilde\epsilon_1-\tilde\epsilon_4-\tilde\epsilon_5+\tilde\epsilon_6)
(-\tilde\epsilon_2+\tilde\epsilon_3+\tilde\epsilon_4-\tilde\epsilon_7)}
\label{eq:se1}
\\
&&
-\bar{f}_1 f_2\bar{f}_3f_4f_5 \bar{f}_6 \bar{f}_7
\frac{e^{-(\tilde\epsilon_1
-\tilde\epsilon_2+\tilde\epsilon_3)\tau}
-e^{-(\tilde\epsilon_1
-\tilde\epsilon_4+\tilde\epsilon_7)\tau}}{
(\tilde\epsilon_1-\tilde\epsilon_4-\tilde\epsilon_5+\tilde\epsilon_6)
(-\tilde\epsilon_2+\tilde\epsilon_3+\tilde\epsilon_4-\tilde\epsilon_7)}.
\label{eq:se2}
\end{eqnarray}
Here (\ref{eq:se1}) corresponds to $S_a$ of (\ref{eq:sa1}) and
(\ref{eq:se2}) to $S_c$ of (\ref{eq:sc}).
Similarly for $S_f$,
\begin{eqnarray}
S_f
&=&
\bar{f}_1 f_2\bar{f}_3f_4f_5 \bar{f}_6 \bar{f}_7
\frac{e^{-(\tilde\epsilon_1
-\tilde\epsilon_4+\tilde\epsilon_7)\tau}
-e^{-(\tilde\epsilon_1
-\tilde\epsilon_2+\tilde\epsilon_3)\tau}}{
(\tilde\epsilon_1-\tilde\epsilon_4-\tilde\epsilon_5+\tilde\epsilon_6)
(-\tilde\epsilon_2+\tilde\epsilon_3+\tilde\epsilon_4-\tilde\epsilon_7)}
\label{eq:sf1}\\
&&
-f_1 f_2\bar{f}_3\bar{f}_4\bar{f}_5 f_6 \bar{f}_7
\frac{e^{-(\tilde\epsilon_5
-\tilde\epsilon_6+\tilde\epsilon_7)\tau}
-e^{-(-\tilde\epsilon_2
+\tilde\epsilon_3+\tilde\epsilon_4+\tilde\epsilon_5-\tilde\epsilon_6)\tau}}{
(\tilde\epsilon_1-\tilde\epsilon_4-\tilde\epsilon_5+\tilde\epsilon_6)
(-\tilde\epsilon_2+\tilde\epsilon_3+\tilde\epsilon_4-\tilde\epsilon_7)}.
\label{eq:sf2}
\end{eqnarray}
At the energy poles at for $\epsilon_\varphi\to i\eta$, $S_f$ becomes identical
to $S_e$. Similarly to the real-time diagrams,
$(-\tilde\epsilon_2+\tilde\epsilon_3+\tilde\epsilon_4-\tilde\epsilon_7)^{-1}$
has a well-defined principal-value integral regardless of the sign
of $\eta$. Therefore for diagrams $S_a-S_f$ we have correct analytic
continuation of imaginary-time results to those of the real-time via
\begin{equation}
\frac{1}{\tilde\epsilon_1-\tilde\epsilon_4-\tilde\epsilon_5+\tilde\epsilon_6}
\to
{\cal
P}\left(\frac{1}{\epsilon_1-\epsilon_4-\epsilon_5+\epsilon_6}\right).
\label{eq:average}
\end{equation}

\begin{figure}
\rotatebox{0}{\resizebox{4.0in}{!}{\includegraphics{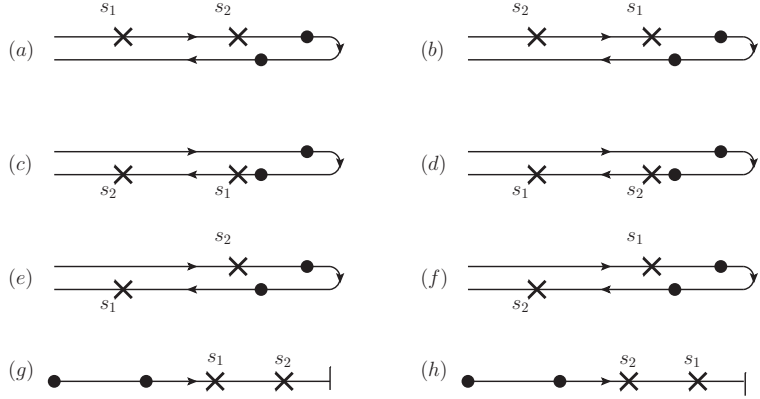}}}
\caption{\small Different time-ordering with two intermediate
interaction events extend to infinity. (a,d,e,g) use the label in
FIG.1(g) and (b,c,f,h) FIG.1(h).
}
\label{fig:fourth2}
\end{figure}

In FIG.~\ref{fig:fourth2}, we consider the remaining time-orderings
with the two intermediate interaction points extending to infinity.
These are harder to deal with, as we discuss below, since the energy
poles may overlap.
\begin{eqnarray}
D_a&=&
f_1 \bar{f}_2f_3\bar{f}_4\bar{f}_5 f_6 \bar{f}_7
\iint\displaylimits_{-\infty}^0 ds_1ds_2 
e^{-i(\epsilon_1
-\epsilon_4-\epsilon_5+\epsilon_6-i\eta)s_1
-i(\epsilon_1-\epsilon_2+\epsilon_3-\epsilon_5+\epsilon_6-\epsilon_7-i\eta)s_2
-i(\epsilon_5-\epsilon_6+\epsilon_7)t}\\
D_b&=&
-\bar{f}_1 f_2\bar{f}_3f_4 f_5 \bar{f}_6 f_7
\int
e^{-i(-\epsilon_1+\epsilon_2
-\epsilon_3+\epsilon_5-\epsilon_6+\epsilon_7-i\eta)s_1
-i(-\epsilon_1+\epsilon_4+\epsilon_5-\epsilon_6-i\eta)s_2
-i(\epsilon_1-\epsilon_2+\epsilon_3)t}\\
D_c&=&
f_1 \bar{f}_2f_3\bar{f}_4\bar{f}_5 f_6 \bar{f}_7
\int
e^{-i(-\epsilon_1+\epsilon_2
-\epsilon_3+\epsilon_5-\epsilon_6+\epsilon_7-i\eta)s_1
-i(-\epsilon_1+\epsilon_4+\epsilon_5-\epsilon_6-i\eta)s_2
-i(\epsilon_5-\epsilon_6+\epsilon_7)t}\\
D_d&=&
-\bar{f}_1 f_2\bar{f}_3f_4 f_5 \bar{f}_6 f_7
\int
e^{-i(\epsilon_1
-\epsilon_4-\epsilon_5+\epsilon_6-i\eta)s_1
-i(\epsilon_1-\epsilon_2+\epsilon_3-\epsilon_5+\epsilon_6-\epsilon_7-i\eta)s_2
-i(\epsilon_1-\epsilon_2+\epsilon_3)t}\\
D_e&=&
-\bar{f}_1 \bar{f}_2f_3f_4f_5 \bar{f}_6 \bar{f}_7
\int
e^{-i(\epsilon_1
-\epsilon_4-\epsilon_5+\epsilon_6-i\eta)s_1
-i(-\epsilon_2+\epsilon_3+\epsilon_4-\epsilon_7-i\eta)s_2
-i(\epsilon_1-\epsilon_4+\epsilon_7)t}\\
D_f&=&
f_1 f_2\bar{f}_3\bar{f}_4 \bar{f}_5 f_6 f_7
\int
e^{-i(\epsilon_2
-\epsilon_3-\epsilon_4+\epsilon_7-i\eta)s_1
-i(-\epsilon_1+\epsilon_4+\epsilon_5-\epsilon_6-i\eta)s_2
-i(-\epsilon_2+\epsilon_3+\epsilon_4+\epsilon_5-\epsilon_6)t}
\end{eqnarray}
After integrals over $s_1$ and $s_2$ it is easy to see that
$D_a(i\eta)=D_c(-i\eta)$ and $D_b(i\eta)=D_d(-i\eta)$.
For $D_e$ and $D_f$, we can swap the dummy indices as
$1\leftrightarrow 7$,
$2\leftrightarrow 6$, and
$3\leftrightarrow 5$, and it becomes 
$D_e(i\eta)=D_e(-i\eta)$ and 
$D_f(i\eta)=D_f(-i\eta)$.
Therefore, we obtain the desired result as (\ref{eq:average}),
\begin{equation}
\sum_{k=a,\cdots,f} D_k(i\eta)=\sum_k D_k(-i\eta)
=\sum_k{\cal P}D_k(\pm i\eta).
\end{equation}
In deriving these relations, no assumptions of $L/R$ and particle-hole
symmetry have been used.
One can rewrite $D_a$ as
\begin{equation}
D_a=
f_1 \bar{f}_2f_3\bar{f}_4\bar{f}_5 f_6
\bar{f}_7\frac{e^{-i(\epsilon_5-\epsilon_6+\epsilon_7)t}}{\epsilon_2-\epsilon_3
-\epsilon_4+\epsilon_7}\left[
\frac{1}{\epsilon_1-\epsilon_2+\epsilon_3-\epsilon_5+\epsilon_6-\epsilon_7-i\eta}
-\frac{1}{\epsilon_1-\epsilon_4-\epsilon_5+\epsilon_6-i\eta}
\right].
\label{eq:da}
\end{equation}
Here the $+i\eta$ in the denominator will be cancelled by $D_c$ and all
fractions can be written as principal-valued, unless the poles
coincide.

We can now turn to the imaginary-time diagrams FIG.~\ref{fig:fourth2}(g,h).
\begin{eqnarray}
D_g
&=&\frac{f_1 \bar{f}_2f_3\bar{f}_4\bar{f}_5 f_6
\bar{f}_7}{-(\tilde\epsilon_2
-\tilde\epsilon_3-\tilde\epsilon_4+\tilde\epsilon_7)}
\left(
-\frac{1}{\tilde\epsilon_1-\tilde\epsilon_2+\tilde\epsilon_3-\tilde\epsilon_5
+\tilde\epsilon_6-\tilde\epsilon_7}
+\frac{1}{\tilde\epsilon_1-\tilde\epsilon_4-\tilde\epsilon_5+\tilde\epsilon_6
}
\right)e^{-(\tilde\epsilon_5-\tilde\epsilon_6+\tilde\epsilon_7)\tau}
\nonumber\\
& & -\frac{\bar{f}_1 f_2\bar{f}_3\bar{f}_4f_5 \bar{f}_6
f_7}{(\tilde\epsilon_2
-\tilde\epsilon_3-\tilde\epsilon_4+\tilde\epsilon_7)}
\frac{e^{-(\tilde\epsilon_1-\tilde\epsilon_2+\tilde\epsilon_3)\tau}}{
(\tilde\epsilon_1-\tilde\epsilon_2+\tilde\epsilon_3-\tilde\epsilon_5
+\tilde\epsilon_6-\tilde\epsilon_7)}
+\frac{\bar{f}_1 \bar{f}_2f_3f_4f_5 \bar{f}_6
\bar{f}_7}{(\tilde\epsilon_2
-\tilde\epsilon_3-\tilde\epsilon_4+\tilde\epsilon_7)}
\frac{e^{-(\tilde\epsilon_1-\tilde\epsilon_4+\tilde\epsilon_7)\tau}}{
(\tilde\epsilon_1-\tilde\epsilon_4-\tilde\epsilon_5
+\tilde\epsilon_6)}.
\end{eqnarray}
After swapping $1\leftrightarrow 7$, $2\leftrightarrow 6$,
and $3\leftrightarrow 5$, the first two terms correspond to $D_a$ and
$D_c$ for $\epsilon_\varphi\to i\eta$ and the third term to $D_e$. Using a similar technique in
(\ref{eq:da}), we can
decouple the product of energy denominators to a sum of simple poles of
$\epsilon_\varphi$ and then by taking the limit Eq.~(\ref{eq:average}),
all energy denominators become
principal-valued, unless poles coincide.

Now we deal with the case when the $\delta$-functions overlap. As
discussed in section~\ref{sec:real}, the double-$\delta$ terms manifest
as terms proportional to $T^2$.
The terms $D_a$, $D_c$ and $D_e$ have double-$\delta$ terms cancelled among
themselves.
At the energy-poles $\epsilon_1-\epsilon_4-\epsilon_5+\epsilon_6=0$ and
$\epsilon_2-\epsilon_3-\epsilon_4+\epsilon_7=0$,
\begin{equation}
D_a=D_c\propto f_1\bar{f}_2 f_3\bar{f}_4\bar{f}_5
f_6\bar{f}_7\frac{T^2}{2}e^{-i(\epsilon_5-\epsilon_6+\epsilon_7)t}.
\end{equation}
For $D_e$, we first rewrite
\begin{equation}
\int_t^Tds_1=\int_0^Tds_1+\int_t^0 ds_1,
\end{equation}
and note that the second integral with a finite interval should not
contribute a $\delta$-function. So as long as double-$\delta$ is concerned,
we only consider the first interval,
\begin{equation}
D_e\propto -\bar{f}_1\bar{f}_2 f_3f_4f_5
\bar{f}_6\bar{f}_7T^2e^{-i(\epsilon_5-\epsilon_6+\epsilon_7)t}
\to -f_1\bar{f}_2 f_3\bar{f}_4\bar{f}_5
f_6\bar{f}_7T^2e^{-i(\epsilon_5-\epsilon_6+\epsilon_7)t}.
\end{equation}
where at the last step the dummy indices are swapped as
$1\leftrightarrow 5$ and $4\leftrightarrow 6$. Therefore, the
double-$\delta$ terms disappear in $D_a+D_c+D_e$. The same is true with
$D_b+D_d+D_f$, and it shows that the all energy poles for the
fourth-order vertex corrections, FIG.~\ref{fig:fourth}(g-h), are
interpreted as principal-valued.

\end{widetext}


\begin{thebibliography}{*}
\bibitem{comment1} Local in this context means that the operator should consist only of creation
and annihilation operators which act on the QD respectively its neighboring sites in the leads.
\end{thebibliography}
\end{document}